\documentclass{article}

\usepackage{algorithm}
\usepackage[margin=30mm,includehead,includefoot]{geometry}
\usepackage[noend]{algpseudocode}
\usepackage{amsmath,amssymb,amsthm,amscd,color,comment}
\usepackage{autobreak}
\usepackage{color}
\usepackage{xcolor}
\usepackage{url}
\usepackage{blkarray}
\usepackage{jheppub}
\usepackage{empheq}
\usepackage{graphicx}
\usepackage{booktabs, multirow}
\usepackage{subcaption}
\usepackage{bm}
\usepackage{enumerate}

\usepackage{tikz}
\usetikzlibrary{patterns}
\usetikzlibrary{arrows,shapes,positioning}
\usetikzlibrary{trees}
\usetikzlibrary{matrix,arrows} 				% For commutative diagram
\usetikzlibrary{positioning}				% For "above of=" commands
\usetikzlibrary{calc,through}				% For coordinates
\usetikzlibrary{decorations.pathreplacing}  % For curly braces
\usepackage{pgffor}							% For repeating patterns
\usetikzlibrary{decorations.pathmorphing}	% For Feynman Diagrams
\usetikzlibrary{decorations.markings}
\tikzstyle{block} = [draw, rectangle, 
minimum height=3em, minimum width=6em]

\tikzstyle{decision} = [diamond, draw, fill=blue!20, 
text width=4.5em, text badly centered, node distance=3cm, inner sep=0pt]
\tikzstyle{block} = [rectangle, draw, fill=blue!10, 
text width=7em, text centered, rounded corners, minimum height=4em, node distance=3cm]
\tikzstyle{line} = [draw, -latex']
\tikzstyle{cloud} = [draw, rectangle, fill=blue!20, text width=6em, text centered, rounded corners, node distance=3cm,
minimum height=2em]
\tikzstyle{border} = [draw, dashed, rectangle, fill=blue!5, rounded corners, node distance=3cm, minimum height=25em, minimum width=22em]
\tikzstyle{data} = [draw, rectangle, fill=blue!10, rounded corners, minimum height=2em, minimum width=8em]
\tikzstyle{box} = [draw, rectangle, fill=white, rounded corners, minimum width=10em]
\tikzstyle{input}=[trapezium, draw, text centered, trapezium left angle=60, trapezium right angle=120, minimum height=2em, fill=blue!10]

\tikzset{fermionnoarrow/.style={draw=black},}

\def\iimg{ {\bf i}}

\definecolor{green1}{HTML}{3D792A}
\definecolor{cyan1}{HTML}{37cdaa}
\definecolor{blue1}{HTML}{5d7ac4}
\definecolor{red1}{HTML}{d0482a}
\definecolor{purple1}{HTML}{845ea8}
\definecolor{orange1}{HTML}{e07229}

\title{Gravitational Spin-Orbit Hamiltonian at NNNLO in the post-Newtonian framework}

\author[a]{Manoj K. Mandal,}
\author[b,a]{Pierpaolo Mastrolia,}
\author[c,d,e]{Raj Patil,}
\author[c]{Jan Steinhoff}

\newcommand{\unipd}{Dipartimento di Fisica e Astronomia, Universit\`a degli Studi di Padova,
Via Marzolo 8, I-35131 Padova, Italy.}

\newcommand{\pdinfn}{INFN, Sezione di Padova,
Via Marzolo 8, I-35131 Padova, Italy.}

\affiliation[a]{\pdinfn}
\affiliation[b]{\unipd}
\affiliation[c]{Max Planck Institute for Gravitational Physics (Albert Einstein Institute), Am M{\"u}hlenberg 1, Potsdam 14476, Germany}

\affiliation[d]{Institut f{\"u}r Physik und IRIS Adlershof, Humboldt-Universit {\"a}t zu Berlin, Zum Großen Windkanal 2, D-12489 Berlin, Germany}
\affiliation[e]{Indian Institute of Science Education and Research Bhopal, Bhopal Bypass Rd, Bhauri, Madhya Pradesh 462066, India.}

\emailAdd{manojkumar.mandal@pd.infn.it}
\emailAdd{pierpaolo.mastrolia@unipd.it}
\emailAdd{raj.patil@aei.mpg.de}
\emailAdd{jan.steinhoff@aei.mpg.de}

\abstract{We present the result of the spin-orbit interaction Hamiltonian for binary systems of rotating compact objects with generic spins, up to \NNNLO corrections within the post-Newtonian expansion. The calculation is performed by employing the effective field theory diagrammatic approach, and it involves Feynman integrals up to three loops, evaluated within the dimensional regularization scheme. 
We apply canonical transformations to eliminate the non-physical divergences and spurious logarithmic behaviours of the Hamiltonian, and use the latter to 
derive the gauge-invariant binding energy and the scattering angle, in special kinematic regimes.
}

\def\NNLO{N$^2$LO }
\def\NNNLO{N$^3$LO }

\allowdisplaybreaks

\begin{document}
\addtocontents{toc}{\protect\setcounter{tocdepth}{2}}

\begin{flushright}
\begingroup\footnotesize\ttfamily
	HU-EP-22/28-RTG
\endgroup
\end{flushright}

\maketitle

\section{Introduction}
The first gravitational wave (GW), emerging from the collision of two black holes, 
which shook the optical systems of the GW observatories \cite{LIGOScientific:2016aoc} can be considered the trigger of a new era in precision astronomy and cosmology. Since then the LIGO-Virgo-KAGRA collaboration detected already 90 GW events \cite{LIGOScientific:2021djp} and the worldwide network of ground-based \cite{LIGOScientific:2014pky,VIRGO:2014yos,KAGRA:2020agh,Saleem:2021iwi,LIGOScientific:2016wof,Punturo:2010zza} as well as space based GW detectors \cite{LISA} continues to grow, granting access to an ever broader frequency band and higher sensitivity.

The experimental advances have been boosting the theoretical investigation, by pushing the application of canonical methods as well as the development of novel techniques aiming at providing the most accurate description of the dynamics of coalescing binary systems.
Earlier developments dedicated to the post-Newtonian (PN)~\cite{Einstein:1938yz, Einstein:1940mt, Ohta:1973je, Jaranowski:1997ky, Damour:1999cr, Blanchet:2000nv, Damour:2001bu, Damour:2014jta, Jaranowski:2015lha, Bernard:2016wrg,Bini:2019nra,Bini:2020wpo,Bini:2020uiq,Bini:2020nsb,Bini:2020hmy,Bini:2020rzn}, 
and the post-Minkowskian (PM) approximation~\cite{Bertotti:1956pxu, Kerr:1959zlt, Bertotti:1960wuq, Portilla:1979xx, Westpfahl:1979gu, Portilla:1980uz, Bel:1981be, Westpfahl:1985tsl, Damour:2016gwp, Damour:2017zjx},
the gravitational self-force formalism~\cite{Mino:1996nk, Quinn:1996am}, 
the effective--one-body formalism~\cite{Buonanno:1998gg,Buonanno:2000ef}, 
as well as the effective field theory (EFT) approach 
for the PN formulation of general relativity
%the non-relativistic general-relativity (NRGR)formalism
~\cite{Goldberger:2004jt,Gilmore:2008gq,Chu:2008xm},
have been flanked by modern Feynman diagrams-
\cite{Foffa:2011ub,Foffa:2012rn,Foffa:2019rdf,Foffa:2016rgu,Foffa:2019yfl,Blumlein:2020pog,Foffa:2019ahl,Foffa:2019hrb, Blumlein:2021txe,Blumlein:2021txj,Blumlein:2020znm}, 
and Scattering Amplitudes-based approaches \cite{Donoghue:1994dn,Bjerrum-Bohr:2002gqz,Neill:2013wsa,Bjerrum-Bohr:2018xdl,Cheung:2018wkq,Kosower:2018adc,Bern:2019crd,Kalin:2019inp, Kalin:2019rwq,Kalin:2020fhe, Kalin:2020mvi,Bern:2021dqo,Bern:2021yeh,Bjerrum-Bohr:2021din,Kol:2021jjc,Brandhuber:2021eyq,DiVecchia:2021bdo,Dlapa:2021npj,Dlapa:2021vgp}, 
which turned out to be particularly powerful for the determination of the perturbative corrections to the classical potential within the EFT approach.

Several studies were also performed to include the spinning motion of the compact objects in the analysis. Initial studies extending the classical techniques were done in \cite{Tulczyjew:1959,Damour:1982}, which was later extended by \cite{Tagoshi:2000zg,Faye:2006gx,Damour:2007nc,Hartung:2011te,Hartung:2013dza,Steinhoff:2008zr,Steinhoff:2009ei,Bohe:2012mr,Aoude:2022thd,FebresCordero:2022jts,Bern:2022kto,Liu:2021zxr,Jakobsen:2022fcj}. The spin effects were then also included in the PN formalism of GR in \cite{Porto:2005ac,Porto:2010tr,Levi:2010zu,Levi:2020kvb,Porto:2006bt,Porto:2008tb,Levi:2008nh,Porto:2008jj,Levi:2011eq,Levi:2014sba,Levi:2015ixa,Kim:2021rfj,Levi:2020uwu,Levi:2019kgk,Levi:2015msa,Levi:2014gsa}. More recently, quantum scattering amplitudes involving massive particles of arbitrary spin were used to obtain classical spin corrections to the two-body effective potential \cite{Guevara:2017csg,Vines:2018gqi,Guevara:2018wpp,Chung:2018kqs,Guevara:2019fsj,Chung:2019duq,Siemonsen:2019dsu,Guevara:2020xjx,Arkani-Hamed:2019ymq}.
See \cite{Levi:2018nxp,Porto:2016pyg,Blanchet:2013haa} for recent reviews and a more comprehensive reference to literature.

In particular, the PN analysis follows the system during the so-called {\it inspiral phase}, where
the components of the binary move at non-relativistic velocities and their orbital separation is slowly decaying. At this stage, the non-relativistic regime of the motion allows for a perturbative treatment of the problem, that can be studied by a series expansion in powers of $v/c$, where $v$ is the orbital velocity of the compact binary and $c$ is the speed of light. For bound state systems like binaries of compact objects, in this phase, the kinetic and potential energy must share the same order of magnitude according to the virial theorem, namely $v^2 = G_N M / r$, where $G_N$ is the Newton’s constant, $M$ is the typical mass of the system and $r$ is the separation between the two bodies. From this perspective, the PN approach is an expansion in two parameters $v$ and $G_N$, whose $n^{th}$ order
%, which 
is referred to as $n$PN (or usually referred to as N$^n$LO), has contributions from terms proportional to $G_N^l (v^2)^{n+1-l}$ where $l = 1, 2, ..., n + 1$.
The analysis then is separated in two sectors, the {\it conservative} sector, where the outgoing radiation is ignored and orbits don't decay, and the {\it radiation} sector, where the emitted radiation is analysed. 
In the radiative sector, the emitted radiation carries energy and angular momentum to infinity. The radiations could also scatter off the background curvature to interact with the orbital dynamics giving rise to the so-called {\it tail} effects, which contributes to both the conservative and the radiative sector. 
See \cite{Blanchet:2013haa, Porto:2016pyg, Levi:2018nxp} for reviews.

The EFT approach is very well suited for this problem because it can be well separated into three different length scales, namely, the length scale associated with the compact object $R_s$ (Schwarzschild radius), the radius of the orbit $r$, and the wavelength of the emitted gravitational wave $\lambda$. Then an EFT at each scale exactly does the job of determining the essential degrees of freedom. 
In this method, the determination of any observable quantity at a given PN is equivalent to the computation of corresponding scattering amplitudes, which can be systematically computed through the evaluation of corresponding Feynman diagrams.
The leading contribution to scattering amplitudes is given by tree diagrams, whereas higher order corrections are determined by diagrams with multiple loops. Tree diagrams represent rational functions of the kinematic variables (energy, momenta, and masses of the particles), therefore they are easy to compute. Loop diagrams, instead, represent very challenging integrals. This means that computing a scattering amplitude for each new level of precision requires drawing exponentially more Feynman diagrams and solving a vastly more complicated mathematical formula. 

In this article we focus on the EFT techniques developed in \cite{Goldberger:2004jt,Porto:2005ac} to carry the post-Newtonian analysis in the conservative sector of the two body problem.

The analysis of compact objects without spin up to the 1PN correction in the conservative orbital dynamics was reported in
\cite{Goldberger:2004jt} for the first time using EFT techniques. Following this, the
2PN was computed in \cite{Gilmore:2008gq,Chu:2008xm}, the 3PN was computed in \cite{Foffa:2011ub} and the 4PN was computed in \cite{Foffa:2012rn,Foffa:2019rdf,Foffa:2016rgu,Foffa:2019yfl,Blumlein:2020pog}. The current state of the art for the post-Newtonian calculation for the conservative potential is the 5PN correction computed in \cite{Foffa:2019hrb,Blumlein:2021txe}. Several partial results of 6PN \cite{Blumlein:2021txj,Blumlein:2020znm} have been recently added to the catalogue.

Compact objects are expected to be rotating rapidly and hence the analysis of the spin becomes important for systems with such objects. The analysis of spin in the formalism of EFTs was initiated in \cite{Porto:2005ac}. For systems with spinning compact objects, the post-Newtonian analysis is usually separated into different sectors with different powers of the spin variable appearing in the effective Lagrangian. 
The spin-orbit sector is analogous to the fine structure correction obtained to the hydrogen atom, which describe the interaction between the orbital angular momentum of the binary and the spin of one of its constituents. Whereas, the quadratic in spin are analogous to the hyperfine structure corrections to the hydrogen atom and describe the interaction between each other/self of the spins of the binary constituents. 

For the spin-orbit sector, the leading order (LO) effective potential was first computed in \cite{Porto:2005ac}. The next-to-leading order (NLO) potential in \cite{Porto:2010tr,Levi:2010zu}, and \NNLO in \cite{Levi:2015uxa}. Partial results for the \NNNLO correction were recently presented in \cite{Levi:2020kvb}.
Similarly, for the quadratic in spin sector, the LO effective potential was also computed in \cite{Porto:2005ac}. The NLO in \cite{Porto:2006bt,Porto:2008tb,Levi:2008nh,Porto:2008jj}, 
\NNLO in \cite{Levi:2011eq,Levi:2014sba,Levi:2015ixa} and \NNNLO was computed in \cite{Kim:2021rfj,Levi:2020uwu}.
The computation for cubic and higher orders in the spin variables can be found in \cite{Levi:2019kgk,Levi:2014gsa,Levi:2015msa}, whereas the finite size effects are described in detail in  \cite{Porto:2008jj,Levi:2014gsa,Levi:2015msa}.
The spin-orbit \NNNLO was then completed in \cite{Antonelli:2020aeb} using the techniques of first-order self-force. 
Here, for maximally rotating compact objects, the N$^n$LO spin-orbit sector contributes at $(3/2+n)$PN order and N$^n$LO spin-squared sector contributes at $(2+n)$PN order.

In this article, we present the complete conservative \NNNLO spin-orbit interaction,
using the framework of EFT~\cite{Goldberger:2004jt,Porto:2005ac} and an extension of the diagrammatic approach  presented in~\cite{Foffa:2016rgu}. In particular,
we begin with deriving the required Feynman rules and then the Feynman diagrams:
 $1 \, (0\text{PN}) + 4 \, (1\text{PN}) + 21 \, (2\text{PN}) + 130 \, (3\text{PN})$ = 156, for the non-spinning sector;
 $ 2 \, (\text{LO}) + 13 \, (\text{NLO}) + 100 \, (\text{\NNLO}) + 894 \, (\text{\NNNLO}) $ = 1009, for the spinning sector,
The corresponding tensor integrals are decomposed in terms of scalar loop integrals, up to three loops, which are  decomposed in terms of a minimal basis of master integrals (MIs), by means of integration-by-parts (IBP) identities \cite{Chetyrkin:1981qh,Laporta:2000dsw}. The contribution of each diagram to the scattering amplitude is then obtained after substituting the analytic expression of the MIs. Finally,
the Fourier transform of the amplitude generates the the effective Lagrangian, later converted into a Hamiltonian which is the main result of this article. 
The computational setup has been fully automated within an in-house 
\texttt{Mathematica} routine
interfaced to 
\texttt{QGRAF}~\cite{NOGUEIRA1993279}, used for the diagram generation,
\texttt{xTensor}~\cite{xAct}, for tensor algebra manipulation,
\texttt{LiteRed}~\cite{Lee:2013mka}, for the IBP decomposition, 
elaborating on some of the ideas implemented in \texttt{EFTofPNG}~\cite{Levi:2017kzq}.

The effective Lagrangian obtained in this way contain higher order derivatives of the position and the spin, which are eliminated by employing suitable coordinate transformations. The EFT Hamiltonian is then obtained by applying the Legendre transform, and it is found to contains poles in $\epsilon=d-3$ 
($d$ being the number of the continuous space dimensions), and logarithmic terms depending on the size of the binary system, which are eliminated by suitable canonical transformations.

Our novel result for the spin-orbit interaction Hamiltonian are used to derive the binding energy for circular orbits and the scattering angle, and we are glad to report that both expressions agree with the results given in \cite{Antonelli:2020aeb}.

The paper is organised as follows. In section \ref{sec_setup}, we review
the description of the spinning binaries within the EFT formalism.  In section
\ref{sec_Routine}, we present the computation for the N$^3$LO spin-orbit potential 
employing the Feynman diagrammatic approach within the EFT framework. 
Then, in section
\ref{sec_Processing_the_effective_potential}, we describe the procedure of removing the
residual divergences and logarithms to derive the EFT Hamiltonian.
We provide our main result of the spin-orbit Hamiltonian up to \NNNLO in section 
\ref{sec_results}. In section \ref{sec_compputing_observables}, we compute two 
observable from the EFT Hamiltonian, namely, the binding energy of the binary system in 
circular orbits with aligned spin configuration and scattering angle for two spinning 
compact objects with aligned spins. We summarize our main results
in section \ref{sec_Conclusion}.
The article contain three appendices: in appendix \ref{app_notation_and_convention}, we describe the notations used in this
article; in appendix \ref{app_master_and_fourier_ints}, we provide the required 
master integrals and their expression; in appendix~\ref{app_Ham}, 
we provide the Hamiltonians till 3PN in the non-spinning sector and till NNLO in the spin-orbit sector.
\\ \\
We provide the required EFT Feynman rules 
in the ancillary file \texttt{Feynman\_Rules.m} and the analytic results of the Hamiltonian till \NNNLO in the ancillary file \texttt{Hamiltonian.m}.
\section{EFT of spinning objects}\label{sec_setup}

In this section, we describe the action for the degrees of freedom of the gravitational field and the degrees of freedom of the compact objects, namely their center of mass and their spin. Then we describe the techniques of the Post-Newtonian formulation of GR in detail and briefly outline the procedure to compute the effective action.

\subsection{Action}

The dynamics of the gravitational field ($g_{\mu\nu}$) is given by the Einstein-Hilbert action along with a gauge fixing term,
\begin{align}\label{eq_action_EH}
S_{\text{EH}} = -\frac{c^4}{16 \pi G_N} \int d^4x \sqrt{g} ~R[g_{\mu\nu}] + \frac{c^4}{32 \pi G_N } \int d^4x \sqrt{g} ~g_{\mu\nu}\Gamma^\mu \Gamma^\nu \, ,
\end{align}
where $\Gamma^\mu=\Gamma^{\mu}_{\rho\sigma}g^{\rho\sigma}$ (in the harmonic gauge $\Gamma^\mu = 0$), $\Gamma^{\mu}_{\rho\sigma}$ is the Christoffel symbol, $G_N$ is the Newton's constant,  $R$ is the Ricci scalar, and $g$ is the determinant of the $g_{\mu\nu}$. 

To model the dynamics of the compact object, we utilize worldlines $x_{(a)}^\mu(\tau)$ parametrized by an affine parameter $\tau$ and define tetrads $\Lambda^\mu_{(a)A}(\tau)$ along the worldlines which connects the body-fixed frame (denoted by upper case Latin indices) of the $a^{\text{th}}$ compact object and the general coordinate frame (denoted by Greek indices). Then the angular velocity tensor of the spinning object can be defined as
\begin{align}
\Omega_{(a)}^{\mu\nu}=\Lambda^\mu_{(a)A} \frac{d\Lambda_{(a)}^{A\nu}}{d\tau} \, ,
\end{align} 
and the corresponding conjugate momenta to the $\Lambda^\mu_{(a)A}$ is given by
\begin{align}
S_{(a)\mu\nu}=-2\frac{\partial L_{\text{pp}}}{\partial \Omega_{(a)}^{\mu\nu}}\, .
\end{align}
Then by demanding the reparameterization invariance of the point particle action, the dynamics for the compact objects is governed by the worldline point particle action given by \cite{Levi:2015msa},
\begin{align}\label{eq_action_pp}
S_{\text{pp}} = \sum_{a=1,2} \int d\tau \left( - m_{(a)}c \sqrt{u_{(a)}^2} -\frac{1}{2} S_{(a)\mu\nu} \Omega_{(a)}^{\mu\nu} - \frac{S_{(a)\mu\nu}u_{(a)}^\nu}{u_{(a)}^2}\frac{du_{(a)}^\mu}{d\tau} +\mathcal{L}_{(a)\text{SI}} \right)\, ,
\end{align}
which, corresponds to Pryce, Newton, and Wigner gauge for spin supplementarity condition (SSC) given by $S_{(a)\mu\nu} (u_{(a)}^\nu + \sqrt{u_{(a)}^2} \delta^{\nu 0})\approx0$.
In the above action, the center of mass of the compact object is modeled by the position of the point particle $x_{(a)}^\mu$ and the spin of the compact object is modeled by $S_{(a)\mu\nu}$. The $\mathcal{L}_{(a)\text{SI}}$ is given by equation (4.16) of \cite{Levi:2015msa} which corresponds to the spin-induced non minimal couplings that does not contribute up to the \NNNLO spin-obit sector at 4.5PN.  
Here $u_{(a)}^\mu=dx_{(a)}^\mu/d\tau$ is the four-velocity and $u_{(a)}^2=g_{\mu\nu}u_{(a)}^\mu u_{(a)}^\nu$, whereas the proper time $\tau$ is related to the coordinate time $t$ by $d\tau=c~ dt$.

As we will be performing the computation using the techniques of multi-loop Feynman diagrams, it is necessary to write the gravitational coupling constant in $d$ dimensions as
\begin{align}
G_d = G_N \Big(\sqrt{4\pi e^{\gamma_E}} R_0\Big)^{d-3}\, ,
\end{align}
where, $R_0$ is an arbitrary length scale.

\subsection{Post-Newtonian formulation of General Relativity}

In the bound state of two compact objects, we have three length scales, namely the length scale associated with the compact object $R_s$ (Schwarzschild radius), the radius of the orbit $r$, and the wavelength of the emitted gravitational wave $\lambda$. We assume the velocities of the particles to be small as compared to the velocity of light and the particles are far from each other, hence propagate on a flat
background ($g_{\mu\nu}=\eta_{\mu\nu}+h_{\mu\nu}$), where the gravitational interaction between the two particles is governed by the gravitons $h_{\mu\nu}$. Then we have a hierarchy of length scales
\begin{align}
\lambda\gg r\gg R_s\, .
\end{align}
As we are only interested in the long-distance physics at the scales of $\lambda$, we first decompose the graviton fields in short distance modes - potential gravitons $H_{\mu\nu}$ with scaling $(k_0,\textbf{k})\sim(v/r,1/r)$ and long-distance modes - radiation gravitons $\bar{h}_{\mu\nu}$ with scaling $(k_0,\textbf{k})\sim(v/r,v/r)$ \cite{Goldberger:2004jt}.

Noting that $v^2 \sim 1/r$ for bound orbits due to the virial theorem (or the third Kepler law), the dimensionless expansion parameter can be taken as $v^2 / c^2 \sim G_N M / r c^2$, which formally scales as $1/c^2$.
Hence, following the majority of the PN literature, we equivalently adopt a formal expansion in $1/c$ with one PN order corresponding to $1/c^2$.
For the spin variables, it holds $S_{(a)} = G m_{(a)}^2 \chi_{(a)} / c$ where the dimensionless spins $\chi_{(a)}$ are at most $\mathcal{O}(1)$ for black holes and (realistic) neutron stars, so that $S_{(a)} \sim 1/c$.
Henceforth we rescale the spins as $S_{(a)} \rightarrow S_{(a)} / c$ in order to make the PN counting in $1/c$ manifest.

Now to compute the conservative binding potential of the two-body system, we ignore the radiation modes and decompose the potential modes in the Kaluza-Klein (KK) parameterization \cite{Kol:2007bc,Kol:2007rx}. In this, the different components of metric $g_{\mu\nu}$ ($=\eta_{\mu\nu}+H_{\mu\nu}$) are encoded in three fields, a scalar $\bm{\phi}$, a 3-dimensional vector $\bm{A}_i$ and a 3-dimensional symmetric rank two tensor $\bm{\sigma}_{ij}$. The decomposition is given by, 
\begin{equation}
g_{\mu\nu} = 
\begin{pmatrix}
e^{2\bm{\phi}/c^2} \,\,\, & -e^{2\bm{\phi}/c^2} \frac{\bm{A}_j}{c^2}\\
-e^{2\bm{\phi}/c^2} \frac{\bm{A}_i}{c^2} \,\,\,\,\,\, & -e^{-2\bm{\phi}/((d-2)c^2)}\bm{\gamma}_{ij}+e^{2\bm{\phi}/c^2} \frac{\bm{A}_i}{c^2}\frac{\bm{A}_j}{c^2}  
\end{pmatrix}\, ,
\end{equation}
where, $\bm{\gamma}_{ij}=\bm{\delta}_{ij}+\bm{\sigma}_{ij}/c^2$.

The 2-body effective action is then given by integrating out the gravitational degrees of freedom from the above derived actions as
\begin{align}
\text{exp}\Big[{\iimg \int dt ~ \mathcal{L}_{\text{eff}}}\Big] = \int D\bm{\phi} D\bm{A}_i D\bm{\sigma}_{ij} ~e^{\iimg(S_{\text{EH}}+S_{\text{pp}})} \, ,
\end{align}
where, $\mathcal{L}_{\rm eff}$ is the effective Lagrangian further decomposed as 
\begin{equation}
 \mathcal{L}_{\rm eff} = \mathcal{K}_{\rm eff} - \mathcal{V}_{\rm eff}\, ,
\end{equation}
where, $\mathcal{K}_{\rm eff}$ is the kinetic term and $\mathcal{V}_{\rm eff}$ is the effective contribution due to gravitational interactions between the two objects.

The effective potential can be represented in the form of connected, classical, 1 particle irreducible (1PI) scattering amplitudes as
\begin{align}
\mathcal{V}_{\text{eff}}=\iimg \lim_{d\rightarrow 3} \int_\textbf{p} e^{\iimg \textbf{p}\cdot (\textbf{x}_{(1)}-\textbf{x}_{(2)})}\quad 	
%\mathcal{V}_{\text{eff}}=\iimg \lim_{d\rightarrow 3} \int_p e^{\iimg p\cdot (\textbf{x}_{(1)}-\textbf{x}_{(2)})}\quad
\parbox{25mm}{
\begin{tikzpicture}[line width=1 pt,node distance=0.4 cm and 0.4 cm]
\coordinate[label=left: ] (v1);
\coordinate[right = of v1] (v2);
\coordinate[right = of v2] (v3);
\coordinate[right = of v3] (v4);
\coordinate[right = of v4, label=right: \tiny$(2)$] (v5);
\coordinate[below = of v1] (v6);
\coordinate[below = of v6, label=left: ] (v7);
\coordinate[right = of v7] (v8);
\coordinate[right = of v8] (v9);
\coordinate[right = of v9] (v10);
\coordinate[right = of v10, label=right: \tiny$(1)$] (v11);
\fill[black!25!white] (v8) rectangle (v4);
\draw[fermionnoarrow] (v1) -- (v5);
\draw[fermionnoarrow] (v7) -- (v11);
\end{tikzpicture}
}
\, ,
\label{eq:effective_lagrangian}
\end{align}
where $\textbf{p}$ is the momentum transfer between the two particles and the box diagram in the above equation refers to all possible Feynman diagrams with gravitons ($\bm{\phi}$, $\bm{A}_i$, and $\bm{\sigma}_{ij}$) mediating the gravitational interaction between the two point particle represented by the two solid black lines.

Our aim in this article is to compute the spin-orbit effective potential up to \NNNLO. For this, we further decompose the kinetic and potential terms as $\mathcal{K}_{\rm eff} = \mathcal{K}_{\rm pp} + \mathcal{K}_{\rm spin}$ and  $\mathcal{V}_{\rm eff} = \mathcal{V}_{\rm pp} + \mathcal{V}_{\rm spin}$ where $\mathcal{K}_{\rm pp}$ and $\mathcal{V}_{\rm pp}$ represent the kinetic and potential terms for center of mass degrees of freedom for the point particle and $\mathcal{K}_{\rm spin}$ and $\mathcal{V}_{\rm spin}$ represent the kinetic and potential terms for the spin degrees of freedom. The expression for the kinetic terms are given by
\begin{equation}\label{eq_Kpp}
\mathcal{K}_{\rm pp} = \sum_{a=1,2} m_{(a)} 
\left[ 
\frac{1}{2}  \textbf{v}_a^2 + \frac{1}{8}  \textbf{v}_{(a)}^4 \left(\frac{1}{c^2}\right)  + \frac{1}{16}  \textbf{v}_{(a)}^6 \left(\frac{1}{c^4}\right) + \frac{5}{128}  \textbf{v}_{(a)}^8 \left(\frac{1}{c^6}\right) 
\right]
+ \mathcal{O}\left(\frac{1}{c^8}\right)\, ,
\end{equation}
\begin{align}\label{eq_KSO}
\mathcal{K}_{\rm spin} = \sum_{a=1,2} 
\Bigg\{ -\frac{1}{2} \textbf{S}_{(a)}^{ij} \bm{\Omega}_{(a)}^{ij} \left(\frac{1}{c}\right)+\textbf{S}_{(a)}^{ij}\textbf{v}_{(a)}^i\textbf{a}_{(a)}^j \left(\frac{1}{c^3}\right) &
\Bigg[
\frac{1}{2}+ \frac{3}{8}  \textbf{v}_{(a)}^2 \left(\frac{1}{c^2}\right) + \frac{5}{16}  \textbf{v}_{(a)}^4 \left(\frac{1}{c^4}\right) \nonumber\\
&+ \frac{35}{128}  \textbf{v}_{(a)}^6 \left(\frac{1}{c^6}\right)
\Bigg]
+ \mathcal{O}\left(\frac{1}{c^{11}}\right) \Bigg\}\, ,
\end{align}
 and the decomposition of the potential terms is defined as follows
\begin{equation}
    \mathcal{V}_{\rm pp}= \mathcal{V}_{\rm N} + \left(\frac{1}{c^2}\right) \mathcal{V}_{\rm 1PN} + \left(\frac{1}{c^4}\right) \mathcal{V}_{\rm 2PN} + \left(\frac{1}{c^6}\right) \mathcal{V}_{\rm 3PN} + \mathcal{O}\left(\frac{1}{c^8}\right)\, ,
\end{equation}
\begin{equation}
	\mathcal{V}_{\rm spin}=  \left(\frac{1}{c^3}\right)
	\left[
	\mathcal{V}^{SO}_{\rm LO} + 
	\left(\frac{1}{c^2}\right) \mathcal{V}^{SO}_{\rm NLO} + \left(\frac{1}{c^4}\right) \mathcal{V}^{SO}_{\rm N^2LO} + \left(\frac{1}{c^6}\right) \mathcal{V}^{SO}_{\rm N^3LO}
	\right]
	+ \mathcal{O}\left(\frac{1}{c^{11}}\right)\, ,
\end{equation}
where, $\mathcal{V}_{\rm N}$ stands for the Newtonian potential and $\mathcal{V}_j$ with $j={\{\rm 1PN,\, 2PN,\, 3PN\}}$ refers to the corresponding PN correction for the non-spinning part of the potential. The  $\mathcal{V}^{SO}_j$ with $j=\{ \rm LO,\, NLO,
\, N^2LO,$ $\rm\, N^3LO \}$ refers to the corresponding correction to the spin-orbit coupling of the binary system. Then our aim in this article would be to compute the $\mathcal{V}_j$ and $\mathcal{V}^{SO}_j$ using the techniques of multi-loop scattering amplitude, as further described in the next sections.

\section{Computational Algorithm}
\label{sec_Routine}

To obtain the effective potential from the diagrammatic approach as shown in equation \eqref{eq:effective_lagrangian}, we begin by generating all the relevant generic topologies contributing at different orders of $G_N$. It could be easily seen from the virial theorem that N$^n$LO has contributions from terms proportional to $G_N^l$ where $l = 1, 2, ..., n + 1$, and we consider all the topologies at $l-1$ loops contributing to the specific order $l$.
So, for the computation of the N$^3$LO spin-orbit potential, we generate all the topologies till the order $G_N^4$ (3-loop)
using \texttt{QGRAF}~\cite{NOGUEIRA1993279}. 
There is 1 topology at order $G_N$ (tree-level), 2 topologies at order $G_N^2$ (one-loop), 9 topologies at $G_N^3$ (two-loop), and 32 topologies at order $G_N^4$ (three-loop). 
Then we dress these topologies with the KK field and Feynman rules derived from the action of PN\footnote{The Feynman rules obtained from the actions in equation \eqref{eq_action_EH} and \eqref{eq_action_pp} after the KK parameterization are provided in an ancillary file \texttt{Feynman\_Rules.m} with this article.} expansion of GR
given in \eqref{eq_action_EH} and \eqref{eq_action_pp}, to obtain all the Feynman diagrams
that contribute to the given order of $G_N$ and $v$ depending on the specific perturbation order. 
The number of diagrams that contribute at particular order in $1/c$ and of particular loop topology are given in table \ref{tbl_no_diag_non_spinning} and \ref{tbl_no_diag_SO}\footnote{While considering the spin effects, we count only the representative Feynman diagrams, where the spin can contribute from any of the world-line graviton interaction vertex present in the diagram. 
Additionally, the diagrams, which can be obtained from the change in the label $1\leftrightarrow 2$, are not counted as separate diagrams.}. 

\begin{figure}[H]%{R}{0.45\textwidth}%[H]
	\centering
	\begin{tikzpicture}[line width=1 pt, scale=0.4]
	\begin{scope}[shift={(-7,0)}]
	%\draw[thick, directed](0,0) ellipse (2cm and 1cm);
	\filldraw[color=gray!40, fill=gray!40, thick](0,0) rectangle (3,3);
	\draw (-1.5,0)--(4.5,0);
	\draw (-1.5,3)--(4.5,3);
	\node at (1.5,6.5) {Gravity};	
	\node at (1.5,5.3) {Diagrams};	
	\end{scope}
	\begin{scope}[shift={(0,0)}]
	\node at (0,6) {$\longleftrightarrow$};
	\node at (0,1.5) {$\equiv$};
	\end{scope}
	\begin{scope}[shift={(5,0)}]
	%\draw[thick, directed](0,0) ellipse (2cm and 1cm);
	\filldraw[color=gray!40, fill=gray!40, thick](0,1.5) circle (1.5);
	\draw (0,3)--(0,4);
	\draw (0,0)--(0,-1);	
	\node at (0,6.5) {Multi-loop};				
	\node at (0,5.3) {Diagrams};				
	\end{scope}		
	\end{tikzpicture}
	\caption{%Relation between different diagrams
	}
	\label{fig_relation_bet_diags}
\end{figure}

\begin{table}
\begin{subtable}[H]{0.48\textwidth}
\begin{tabular}{|c|c|c|c|}
\hline
Order                & Diagrams            & Loops       & Diagrams  \\ \hline
0PN                  & 1                   & 0      & 1   \\ \hline
\multirow{2}{*}{1PN} & \multirow{2}{*}{4}  & 1    & 1 \\ \cline{3-4} 
                     &                     & 0  & 3 \\ \hline
\multirow{3}{*}{2PN} & \multirow{3}{*}{21} & 2    & 5 \\ \cline{3-4} 
                     &                     & 1 & 10 \\ \cline{3-4} 
                     &                     & 0  & 6 \\ \hline
\multirow{4}{*}{3PN} & \multirow{4}{*}{130} & 3    & 8 \\ \cline{3-4} 
                     &                     & 2 & 75 \\ \cline{3-4} 
                     &                     & 1  & 38 \\ \cline{3-4}
                     &                     & 0  & 9 \\ \hline
\end{tabular}
\caption{Non-spinning sector}
\label{tbl_no_diag_non_spinning}
\end{subtable}
\begin{subtable}[H]{0.48\textwidth}
\begin{tabular}{|c|c|c|c|}
\hline
Order                & Diagrams            & Loops      & Diagrams  \\ \hline
LO                  & 2                   & 0      & 2   \\ \hline
\multirow{2}{*}{NLO} & \multirow{2}{*}{13}  & 1    & 8 \\ \cline{3-4} 
                     &                     & 0  & 5 \\ \hline
\multirow{3}{*}{\NNLO} & \multirow{3}{*}{100} & 2    & 56 \\ \cline{3-4} 
                     &                     & 1 & 36 \\ \cline{3-4} 
                     &                     & 0  & 8 \\ \hline
\multirow{4}{*}{\NNNLO} & \multirow{4}{*}{894} & 3    & 288 \\ \cline{3-4} 
                     &                     & 2 & 495 \\ \cline{3-4} 
                     &                     & 1  & 100 \\ \cline{3-4}
                     &                     & 0  & 11 \\ \hline
\end{tabular}
\caption{Spin-orbit sector}
\label{tbl_no_diag_SO}
\end{subtable}
\caption{Number of Feynman diagrams contributing different sectors.}
\end{table}

Within the EFT framework, the sources remain static and as a result the generated 
Feynman diagrams are mapped to two-point multi-loop
Feynman diagrams with mass-less internal lines and an external momentum (the momentum transferred between two sources) as shown in figure \ref{fig_relation_bet_diags}.
We translate these Feynman diagrams to their corresponding Feynman amplitudes after performing the tensor algebra using \texttt{xTensor} \cite{xAct}. 
The generic form of the effective potential corresponding to any $l$-loop Feynman graph $G$ can be expressed 
as
\begin{equation}
{\cal V}_{G}^{(l)} = %(x_{(a)},S_{(a)})  = 
\underbrace{ \vphantom{\sum_{\substack{a\\ b}}} N^{\mu_1,\mu_2,\cdots}_{C~~~\nu_1,\nu_2,\cdots}
\big(x_{(a)},\cdots,S_{(a)},\cdots\big)}_{\substack{\textrm{Coefficient that depends} \\ \textrm{on orbital variables}}} ~
\underbrace{ \vphantom{\sum_{\substack{a\\ b}}} \int_p e^{\iimg p_\mu (x_{(1)}-x_{(2)})^\mu} N_{F~~~\mu_1,\mu_2,\cdots}^{\alpha_1,\alpha_2,\cdots} (p)}_\textrm{Fourier integral} ~
\underbrace{ \vphantom{\sum_{\substack{a\\ b}}} \prod_{i=1}^l \int_{k_i}  
\frac{N_{M~~~\alpha_1,\alpha_2,\cdots}^{\nu_1,\nu_2,\cdots} (k_i)}{\prod_{\sigma \in G} D_\sigma (p, k_i)}}_\textrm{Multi-loop integral}\, ,
\end{equation}
where, 
\begin{enumerate}[(i)]
\item 
$N_C$ is a tensor polynomial depending on the world-line coordinates ($x_{(a)}^\mu$), 
the spin tensor ($S_{(a)\mu\nu}$), 
and their higher-order time derivatives,
\item $p$ is the momentum transfer between the sources (Fourier momentum), 
\item $N_F$ is the tensor polynomial built out of momenta $p$,
\item $k_i$ are the loop momentums, 
\item $D_\sigma$ denotes the set of denominators corresponding to the internal lines of $G$,
\item $N_M$ stands for a tensor polynomial built out of external momenta $p$ and loop momenta $k_i$.
\end{enumerate}

We perform the reduction of the multi-loop tensor integrals to scalar integrals by applying a set of projectors 
exploiting the Lorentz invariance.
We build the projectors depending on the Lorentz invariant 
external momentum($p$) and the background metric.
After applying the projector the numerator ($N_M$) of the multi-loop integral is transformed as
\begin{equation}
    N_{M~~~\alpha_1,\alpha_2,\cdots}^{\nu_1,\nu_2,\cdots} (p,k_i)
    \longrightarrow 
    \tilde{N}_{M} (p,k_i) \tilde{N}_{F~~~\alpha_1,\alpha_2,\cdots}^{\nu_1,\nu_2,\cdots}(p) \, ,
\end{equation}
where,
$\tilde{N}_{M}$ is a polynomial depending on the scalar products between the external momentum ($p$) and the loop momentums ($k_i$).
We use projectors up to rank 6 for the complete evaluation of the \NNNLO Spin-Orbit effective potential.
The generated scalar integrals are not all independent and 
there exist linear relations between these integrals, 
which are envisaged via the
Integration-By-Parts (IBP) relations. 
We employ \texttt{LiteRED} \cite{Lee:2013mka} to generate these IBP relations to obtain a smaller set of independent scalar integrals, 
known as Master integrals (MIs). 
For the specific computation of \NNNLO Spin-Orbit effective potential, 
we obtain 1 MI at one-loop, 2 MIs at two-loop, and 3 MIs at three-loop.
These MIs are well known and admit closed analytic expressions in $d$ dimension.
We provide the explicit expressions
in~\ref{sec_masterints} for convenience.

After the evaluation of the multi-loop integral,
we perform the Fourier transform of the tensor polynomials, which have the form 
\begin{equation}
%    {\cal M}_F = 
    \int_p e^{p_\mu (x_{(1)}-x_{(2)})^\mu} N_{F~~~\mu_1,\mu_2,\cdots}^{\alpha_1,\alpha_2,\cdots} (p) \tilde{N}_{F~~~\alpha_1,\alpha_2,\cdots}^{\nu_1,\nu_2,\cdots}(p) ~ f(p)\, ,
\end{equation}
where, $N_F$ and $\tilde{N}_F$ are the tensor polynomial built out 
of Fourier momenta $p$ and $f(p)$ is a function of the 
external momentum arising from the multi-loop integral.
The Fourier integrals up to rank 8 are required, 
which are obtained by iterative differentiation of the expression given in section \ref{sec_fourierints}.
Then, we obtain the effective potential that only depends on the orbital variables $\textbf{x}_{(a)}$, $\textbf{S}_{(a)}$, and their higher-order time derivatives by expanding the complete scattering amplitude as Laurent series in $\epsilon$ around $d=3$.

\begin{figure}[t]%{R}{0.55\textwidth}
	\begin{tikzpicture}[line width=1 pt, scale=0.58]
	\draw[draw=black, fill=red!10, rounded corners] (-3,2.5) rectangle ++(26,-6);
	\node at (6,2.5) [box] (label) {\sffamily\textbf{Integrand generation}};
	\draw[draw=black, fill=red!10, rounded corners] (-3,-5.5) rectangle ++(26,-7.5);
	\node at (6,-5.5) [box] (label) {\sffamily\textbf{Multi-loop methods}};
	\node at (2,0) [block] (GenTopo) {Topologies};
	\node at (6,-2.5) [input] (GenRules) {Feynman rules};
	\node at (10,0) [block] (GenDiag) {Feynman diagrams};
	\path [line] (GenRules) -- ++(0,2.5);
	\path [line] (GenTopo) -- (GenDiag);
	
	\node at (18,0) [block] (GenInt) {Integrands};
	\path [line] (GenDiag) -- (GenInt);

	\node at (2,-9) [block] (TenRed) {Tensor \\reduction% on loop momentas
	};
	\node at (10,-9) [block] (RedMaster) {IBP Reduction to Master integrals};
	\path [line] (GenInt) |- ++(-16,-4.5) -- (TenRed);
	\path [line] (TenRed) -- (RedMaster);

	\node at (14,-6.6) [input] (MIs) {Master integrals};
	\path [line] (MIs) -- ++(0,-2.4); %-| ++(0,3);

	\node at (18,-9) [block] (FourierInts) {Tensor Fourier integrals};
	\path [line] (RedMaster) -- (FourierInts);
	\path [line] (FourierInts) -| ++(0,-2.2);
	
	\node at (10,-16) [block] (EffLag) {Effective Lagrangian};
	\path [line] (18,-12) |- ++(-8,-2) -- (EffLag);
	
	\node at (18,-11.8) [data] (LaurentExp) {Laurent expansion around $d=3$};

	\end{tikzpicture}
	\caption{Flowchart of the computational algorithm}
	\label{fig_flowchart_of_routine}
\end{figure}
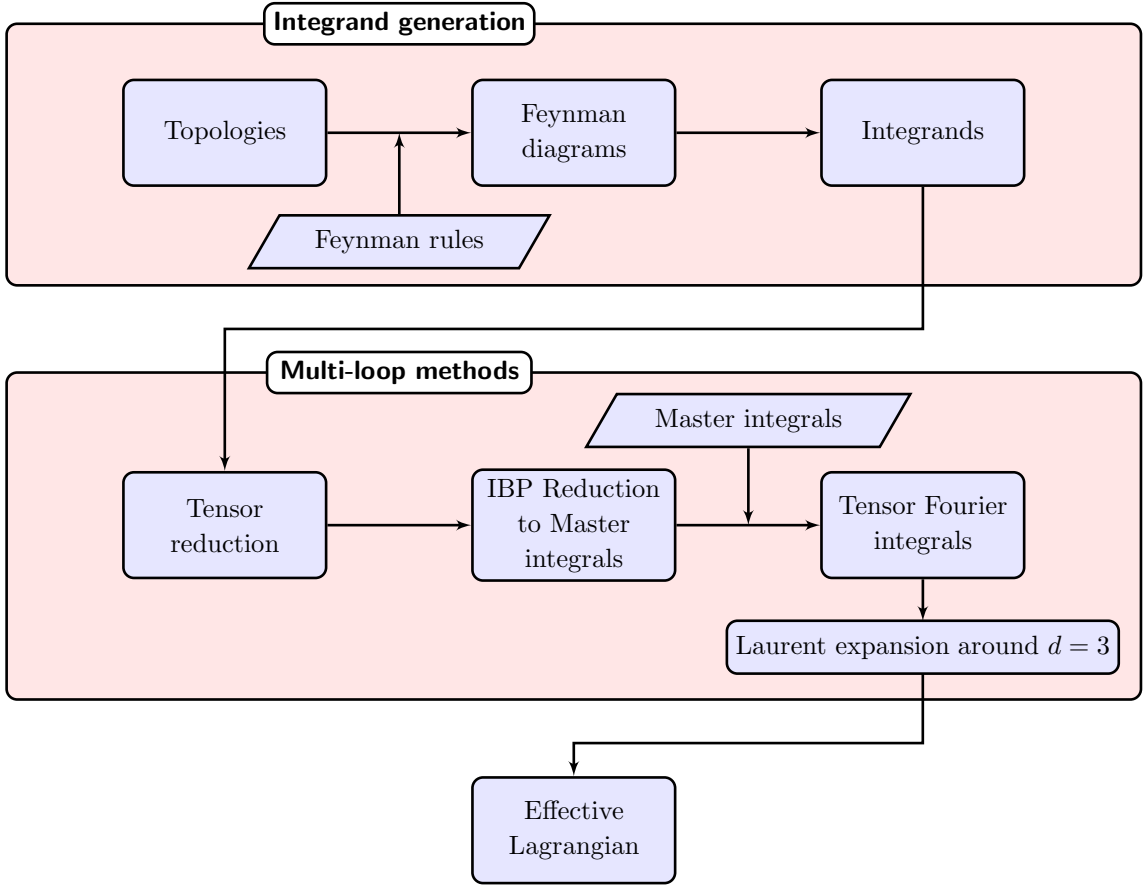

The computation of the effective potential starting from the generation of the required Feynman diagrams, expressing them in multi-loop integrands, performing IBP reduction, and then applying the Fourier transformations have been automated through an in-house code, 
elaborating on some of the ideas implemented in \texttt{EFTofPNG}~\cite{Levi:2017kzq}, 
and using \texttt{xTensor} \cite{xAct} for tensor algebra manipulations as well as successful interface to \texttt{LiteRed}~\cite{Lee:2013mka}, \texttt{Reduze}~\cite{vonManteuffel:2012np}, \texttt{KIRA}~\cite{Maierhofer:2017gsa}  for the IBP reduction. 
A flow chart for the complete computational algorithm for the effective potential as implemented in our in-house code
is shown in Fig. \ref{fig_flowchart_of_routine}.

\section{Processing the effective Lagrangian} \label{sec_Processing_the_effective_potential}

The effective potential obtained in this way usually contains higher-order time derivatives of the position ($\textbf{a}_{(a)}$, $\dot{\textbf{a}}_{(a)}$, $\ddot{\textbf{a}}_{(a)}$,$\cdots$) and the spin ($\dot{\textbf{S}}_{(a)}$, $\ddot{\textbf{S}}_{(a)}$,$\cdots$). 
In our computation of the N$^3$LO spin-orbit potential, we have 6th order time derivative of position and 3rd order of time derivative in spin. 
We need to eliminate these higher-order time derivatives from the potential to facilitate the computation of the Hamiltonian, 
to obtain gauge-invariant quantities as well as to pave the way for the implementation within the effective one-body formalism.
Additionally, the potential in the non-spinning sector at 3PN and 
the N$^3$LO spin-orbit potential contains poles in the dimensional regularization parameter
$\epsilon$ and logarithmic terms of the form log($r \over R_0$). 
These terms may originate either from the choice of coordinates or because of tail effects, and they can be removed, after combining the potential and the tail contribution to conservative effect, by a suitable coordinate transformation \cite{Foffa:2019yfl}. 
However, the tail effects do not arise either upto 3PN order in the non-spinning case, or up to \NNNLO, alias 4.5PN, in spin-orbit sector \cite{Antonelli:2020ybz}, therefore
all the divergent terms and logarithms in our computation 
are removed by finding a proper coordinate transformation.
First, we discuss the procedure to eliminate the higher-order time derivatives and then we show how we remove the divergent pieces  and the logarithmic terms to obtain the EFT Hamiltonian free of any divergences and logarithmic terms.

\subsection{Elimination of higher-order time derivatives}
\label{sec_Elimination_of_higher_order_time_derivatives}

We briefly describe the procedure to remove all the higher-order time derivatives, 
using coordinate transformations~\cite{Levi:2014sba,Schafer:1984mr,Damour:1990jh,Damour:1985mt,Barker:1980spx}. 
The Lagrangian under a coordinate transformation $\textbf{x}_{(a)} \rightarrow \textbf{x}_{(a)} + \delta \textbf{x}_{(a)}$ changes
by
\begin{align}
    \delta \mathcal{L} = %(\textit{EOM for position})^i
    \left( \frac{\delta \mathcal{L}}{\delta \textbf{x}_{(a)}^i} \right)~ \delta \textbf{x}_{(a)}^i + \mathcal{O}\left(\delta \textbf{x}_{(a)}^2\right)
\end{align}
So, when the 
equation of motion
(EOM)
is linear in $\textbf{a}_{(a)}$ at LO,
we can construct a perturbatively small $\delta \textbf{x}_{(a)}$ such that 
terms depending on $\textbf{a}_{(a)}$ drop out in $\mathcal{L}+\delta \mathcal{L}$. 
Similarly, for terms involving higher-order time derivatives of $\textbf{a}_{(a)}$,  one can take $\delta \textbf{x}_{(a)}$ to be a total time derivative such that the higher-order time derivatives in $\mathcal{L}+\delta \mathcal{L}$ cancel upon partial integration.
In general, the $\mathcal{O}\left(\delta \textbf{x}_{(a)}^2\right)$ contributions have to be kept, but will turn out to be negligible for the explicit steps outlined below, making the procedure equivalent to insertion of EOM~\cite{Damour:1990jh}.
Following the same approach, 
when we apply a small transformation simultaneously to the rotation matrix $\bm{\Lambda}_{(a)}^{ij} \rightarrow \bm{\Lambda}_{(a)}^{ij} + \delta\bm{\Lambda}_{(a)}^{ij}$ and the spin $\bm{S}_{(a)}^{ij} \rightarrow \bm{S}_{(a)}^{ij} + \delta\bm{S}_{(a)}^{ij}$.
%which relates the co-rotating frame and the local frame of the spinning object, 
The variation of the rotation matrix, including quadratic terms in $\bm{\omega}_{(a)}$, 
reads as,
\begin{align}
\label{eq:deltaLambda}
\delta\bm{\Lambda}_{(a)}^{ij} = \bm{\Lambda}_{(a)}^{ik} \bm{\omega}_{(a)}^{kj} 
+\frac{1}{2} \bm{\Lambda}_{(a)}^{ik} \bm{\omega}_{(a)}^{kl} \bm{\omega}_{(a)}^{lj} 
+ \mathcal{O}\left(\bm{\omega}_{(a)}^3\right) \ ,
\end{align}
where the $\bm{\omega}_{(a)}^{kj}$ is the antisymmetric generator of the rotation matrix,\footnote{A generic rotation can be written as a matrix exponential $e^{\bm{\omega}_{(a)}}$ such that $\bm{\Lambda}_{(a)} \rightarrow \bm{\Lambda}_{(a)} e^{\bm{\omega}_{(a)}}$ and $\bm{S}_{(a)} \rightarrow e^{-\bm{\omega}_{(a)}} \bm{S}_{(a)} e^{\bm{\omega}_{(a)}}$.} and similarly the spin transforms as
\begin{align}
\delta\bm{S}_{(a)}^{ij} = 2 \bm{S}_{(a)}^{k[i} \bm{\omega}_{(a)}^{j]k} 
+ \bm{S}_{(a)}^{kl} \bm{\omega}_{(a)}^{ik} \bm{\omega}_{(a)}^{jl} + \bm{S}_{(a)}^{k[i} \bm{\omega}_{(a)}^{j]l} \bm{\omega}_{(a)}^{kl}
+ \mathcal{O}\left(\bm{\omega}_{(a)}^3\right) \ .
\end{align}
Due to such transformation, 
the 
Lagrangian
changes by\footnote{Here $V\equiv -\left(\mathcal{L}-\left(-\frac{1}{2} \textbf{S}_{(a)}^{ij} \bm{\Omega}_{(a)}^{ij}\right)\right)$.}
\begin{align}
    \delta \mathcal{L} = &%(\textit{EOM for spin})^{ij}~\bm{\omega}_{(a)}^{ij} + \mathcal{O}\left(\bm{\omega}_{(a)}^2\right)
    - \left(\frac{1}{c}\right) \frac{1}{2} \dot{\bm{S}}_{(a)}^{ij} \bm{\omega}_{(a)}^{ij} - \left(\frac{1}{c}\right) \frac{1}{2} \bm{S}_{(a)}^{ij} \dot{\bm{\omega}}_{(a)}^{ik} \bm{\omega}_{(a)}^{kj}  - \left(\frac{\delta V }{\delta \bm{S}_{(a)}^{ij} }\right) \delta\bm{S}_{(a)}^{ij}  + \mathcal{O}\left(\bm{\omega}_{(a)}^3, \delta\bm{S}_{(a)}^2\right)
\end{align}
So, when the above equation is linear in $\dot{\textbf{S}}_{(a)}$ at LO, 
we can construct the $\bm{\omega}_{(a)}^{ij}$ such that 
all the terms depending on $\dot{\textbf{S}}_{(a)}$
and higher-order derivatives drops out in $\mathcal{L}+\delta \mathcal{L}$. 
Here, it is important to keep terms quadratic in $\bm{\omega}_{(a)}$, as first pointed out in~\cite{Kim:2022pou}, merely inserting EOM would be missing these contributions.
We apply this procedure iteratively to eliminate the higher-order time derivatives from the \NNNLO spin-orbit potential. 
Specifically, we perform 4 iterations, where
\begin{enumerate}
    \item we remove the terms with $\textbf{a}_{(a)}$ and its higher-order time derivatives from the LO and NLO spin-orbit potentials,
    \item we remove  $\dot{\textbf{S}}_{(a)}$ and its higher-order time derivatives from the NLO spin-orbit potential,
    \item we remove the $\textbf{a}_{(a)}$ and its higher-order time derivatives from non-spinning 2PN and 3PN  potentials,
    \item we remove the $\textbf{a}_{(a)}$, $\dot{\textbf{S}}_{(a)}$ and their higher-order time derivatives from the \NNLO and \NNNLO spin-orbit potentials.
\end{enumerate}
After each iteration, we obtain a new Lagrangian (with new EOM) to be used in the next iteration.
One can check that at each step contributions quadratic in $\delta\textbf{x}_{(a)}$ and cubic in $\bm{\omega}_{(a)}$ are negligible (higher order in spin or the PN approximation).
Following these steps, we obtain the effective Lagrangian, which depends on the position, velocity, and spin only.

\subsection{Computation of the EFT Hamiltonians}

We apply the Legendre transformation on the effective Lagrangian obtained 
in the previous step to derive the EFT Hamiltonian 
\begin{align}\label{eq_ham_def}
\mathcal{H}(\textbf{x},\textbf{p},\textbf{S})= \sum_{a=1,2}\textbf{p}_{(a)}^i \dot{\textbf{x}}_{(a)}^i - \mathcal{L}(\textbf{x},\dot{\textbf{x}},\textbf{S}) \ ,
\end{align}
where the canonical momenta $\textbf{p}^i$ is defined as
\begin{align}\label{eq_conjmom_def}
\textbf{p}_{(a)}^i=\frac{\partial \mathcal{L}(\textbf{x},\dot{\textbf{x}},\textbf{S})}{\partial \dot{\textbf{x}}_{(a)}^i}\, .
\end{align}
We invert this relation (order by order in $1/c$) to express $\dot{\textbf{x}}_{(a)}^i$ in terms of $\textbf{p}_{(a)}^i$. We 
use this relation between $\dot{\textbf{x}}_{(a)}^i$ and 
$\textbf{p}_{(a)}^i$ in the equation \eqref{eq_ham_def} to obtain the required Hamiltonian $\mathcal{H}(\textbf{x},\textbf{p},\textbf{S})$.

\subsection{Removal of the poles and logarithms}

The Hamiltonian derived in the previous step, 
and the Lagrangian both
contain divergent pieces and logarithmic terms, 
which can be removed by finding suitable coordinate transformations. 
While using the Lagrangian description, we can add total time derivative terms with arbitrary coefficients, 
and fix these coefficients to a set of values such that the divergent piece drops out while removing the higher-order time derivatives as described in section \ref{sec_Elimination_of_higher_order_time_derivatives}. 

On the other hand, in the Hamiltonian description, we can define a canonical transformation given by
\footnote{The Poisson bracket is defined as
\begin{equation}
    \{A,B\} = \sum_{i=1}^N\left(\frac{\partial A}{\partial r_{(i)}}\frac{\partial B}{\partial p_{(i)}}-\frac{\partial A}{\partial p_{(i)}}\frac{\partial B}{\partial r_{(i)}}\right) +\sum_{i=1}^N\left( \textbf{S}_{(i)} \times \frac{\partial A}{\partial \textbf{S}_{(i)}} \cdot \frac{\partial B}{\partial \textbf{S}_{(i)}} \right) \, ,
\end{equation}
where the last term is ignored in the context of the current analysis since it contributes to higher PN orders.}
\begin{align}
\mathcal{H}'=\mathcal{H}+\{\mathcal{H},\mathcal{G}\} \ ,
\label{eq:Hamiltonian_Cannonical_Transformation}
\end{align}
where $\mathcal{G}$ is the infinitesimal generator of the canonical transformation.
Usually, the ansatz for the terms in the generic total time derivative to be added to the Lagrangian contains a large number of terms.

Here, we remove the poles and the logarithms in the Hamiltonian description by defining a suitable canonical transformation.
We put up an ansatz for the infinitesimal generator $\mathcal{G}$ with arbitrary coefficients and
build a system of linear equations in these coefficients by demanding the 
cancellation of the divergent pieces in $\mathcal{H}'$.
We use the solution to fix the arbitrary coefficients to a set of values, thereby obtaining the final Hamiltonian $\mathcal{H}'$ free of poles and logarithms.
In the following sections, we describe the procedures for the removal of the poles and logarithms from the 3PN non-spinning sector and the Spin-orbit \NNNLO sector.
\subsubsection{3PN non-spinning sector}
The 3PN non-spinning Lagrangian has been studied extensively~\cite{Foffa:2011ub} and we follow the steps suggested to remove the divergent pieces.
Specifically, following~\cite{Foffa:2011ub} we add a total derivative term with the complete 3PN Lagrangian 
\begin{align}
\mathcal{L}_{\text{TD}}=\left(\frac{1}{c^6}\right)\frac{1}{\epsilon}\frac{d}{dt}\left[\frac{G_N^3}{r}\Big(c_1 \left(\textbf{v}_{(1)} \cdot \textbf{n}\right) + c_2 \left(\textbf{v}_{(2)} \cdot \textbf{n}\right)\Big)\right]\, ,
\end{align}
where,
\begin{align}
c_1= \frac{1}{3} \left(4  m_{(1)}^3 m_{(2)}- m_{(1)} m_{(2)}^3\right)\quad\quad,\quad\quad c_2= \frac{1}{3} \left( m_{(1)}^3 m_{(2)}-4  m_{(1)} m_{(2)}^3\right)\, ,
\end{align}
and 
$\textbf{n}\equiv\textbf{r}/r$ is the separation unit vector for the binary.
We follow the usual procedure of removing $\textbf{a}^i$ and $\dot{\textbf{S}}^{ij}$ and their higher-order time derivatives as described in Sec.~\ref{sec_Elimination_of_higher_order_time_derivatives} including the total derivative term with the 3PN Lagrangian.
It turns out that the divergent pieces drop out and we obtain a finite Lagrangian till 3PN. 
Then we remove the terms involving logarithms by finding a canonical transformation to the Hamiltonian obtained from the above Lagrangian. For this purpose, we use the following ansatz for the infinitesimal generator,
\begin{align}
\mathcal{G}_{\text{3PN}}=&\left(\frac{1}{c^6}\right)\frac{G_N^3}{r^2}\left( g_1  \frac{1}{m_{(1)}} \left(\textbf{p}_{(1)} \cdot \textbf{n}\right) + g_2 \frac{1}{m_{(2)}} \left(\textbf{p}_{(2)} \cdot \textbf{n}\right)  \right)\, ,
\end{align}
where,
\begin{align}
g_n \equiv g_{Ln} \log\left(\frac{r}{R_0}\right)\, \, , \, \, \text{for} \,\, n=1,2 \, .
\end{align}
We use this generator to build a canonical transformation following Eq.~\ref{eq:Hamiltonian_Cannonical_Transformation}. By requiring the removal of the logarithm terms in the new Hamiltonian, we build a system of equations and solving them we obtain
\begin{align}
g_{L1}= m_{(1)} m_{(2)}^3-4 m_{(1)}^3 m_{(2)} \quad\quad,\quad\quad g_{L2}= 4 m_{(1)} m_{(2)}^3-m_{(1)}^3 m_{(2)}\, .
\end{align}

\subsubsection{Spin-orbit NNNLO sector}
For the Spin-orbit Hamiltonian at \NNNLO, we remove both the divergent terms as well as terms involving logarithms by finding a suitable canonical transformation. Inspired by~\cite{Levi:2014sba}, we build the following ansatz for the infinitesimal
generator
\begin{align}
\mathcal{G}_{\rm SO-N^3LO}=\left(\frac{1}{c^9}\right)&\Bigg\{\frac{G_N^3}{r^3}\left[g_3  \frac{1}{m_{(1)}m_{(2)}} \Big(\textbf{S}_{(1)}\cdot(\textbf{p}_{(1)}\times\textbf{p}_{(2)})\Big)  + g_4 \frac{1}{m_{(1)}m_{(2)}} \Big(\textbf{S}_{(2)}\cdot(\textbf{p}_{(1)}\times\textbf{p}_{(2)})\Big)\right]\nonumber\\
&+\frac{G_N^3}{r^3}\Bigg[ 
\frac{1}{m_{(1)}} \Big(\textbf{S}_{(1)}\cdot (\textbf{p}_{(1)}\times \textbf{n})\Big) \left(g_{5} \frac{1}{m_{(1)}} \left(\textbf{p}_{(1)} \cdot \textbf{n}\right) + g_{6} \frac{1}{m_{(2)}} \left(\textbf{p}_{(2)} \cdot \textbf{n}\right)\right)\nonumber\\
&\quad\quad\quad +\frac{1}{m_{(2)}} \Big(\textbf{S}_{(1)}\cdot (\textbf{p}_{(2)}\times \textbf{n})\Big) \left(g_{7} \frac{1}{m_{(1)}} \left(\textbf{p}_{(1)} \cdot \textbf{n}\right) + g_{8} \frac{1}{m_{(2)}} \left(\textbf{p}_{(2)} \cdot \textbf{n}\right)\right)\nonumber\\
&\quad\quad\quad +\frac{1}{m_{(1)}} \Big(\textbf{S}_{(2)}\cdot (\textbf{p}_{(1)}\times \textbf{n})\Big) \left(g_{9} \frac{1}{m_{(1)}} \left(\textbf{p}_{(1)} \cdot \textbf{n}\right) + g_{10} \frac{1}{m_{(2)}} \left(\textbf{p}_{(2)} \cdot \textbf{n}\right)\right)\nonumber\\
&\quad\quad\quad +\frac{1}{m_{(2)}} \Big(\textbf{S}_{(2)}\cdot (\textbf{p}_{(2)}\times \textbf{n})\Big) \left(g_{11} \frac{1}{m_{(1)}} \left(\textbf{p}_{(1)} \cdot \textbf{n}\right) + g_{12} \frac{1}{m_{(2)}} \left(\textbf{p}_{(2)} \cdot \textbf{n}\right)\right) \Bigg] \Bigg\}\, ,
\end{align}
where the arbitrary coefficients $g_n$ are defined as
\begin{align}
g_n \equiv g_{\epsilon n} \frac{1}{\epsilon} + g_{Ln} \log\left(\frac{r}{R_0}\right)\, , \, \, \text{for} \,\, n=3,4,\cdots 12 \, .
\end{align}
Following Eqn.~\ref{eq:Hamiltonian_Cannonical_Transformation}, we define a canonical transformation using the above ansatz for the infinitesimal generator and build a system of linear equations by requiring the removal of the divergent as well as the logarithmic terms. We solve these set of equations and find the arbitrary coefficients as 

\begin{align}
&g_{\epsilon 3} = \frac{7 \left(51 m_{(1)}^2 m_{(2)}-25 m_{(2)}^3\right)}{30  } \, , && g_{\epsilon 4}= \frac{7 \left(25 m_{(1)}^3-51 m_{(1)} m_{(2)}^2\right)}{30  }\nonumber\, ,\\
&g_{\epsilon 5}= -\frac{63 m_{(1)}^2 m_{(2)}}{10  } \, , && g_{\epsilon 6}= -\frac{20 m_{(2)}^3-47 m_{(1)}^2 m_{(2)}}{2  }\nonumber\, ,\\
&g_{\epsilon 7}= -\frac{122 m_{(1)}^2 m_{(2)}-75 m_{(2)}^3}{10  } \, , && g_{\epsilon 8}= -\frac{10 m_{(1)}^2 m_{(2)}-77 m_{(2)}^3}{10  }\nonumber\, ,\\
&g_{\epsilon 9}= -\frac{10 m_{(1)} m_{(2)}^2-77 m_{(1)}^3}{10  } \, , && g_{\epsilon 10}= -\frac{122 m_{(1)} m_{(2)}^2-75 m_{(1)}^3}{10  }\nonumber\, ,\\
&g_{\epsilon 11}= -\frac{20 m_{(1)}^3-47 m_{(1)} m_{(2)}^2}{2  } \, , && g_{\epsilon 12}= -\frac{63 m_{(1)} m_{(2)}^2}{10}\nonumber\, ,\\
&g_{L3}= \frac{1}{30} \left(525 m_{(2)}^3-1126 m_{(1)}^2 m_{(2)}\right) \, , &&g_{L4}= \frac{1}{30} \left(1126 m_{(1)} m_{(2)}^2-525 m_{(1)}^3\right)\nonumber\, ,\\
&g_{L5}= \frac{122}{5} m_{(1)}^2 m_{(2)}\, , &&g_{L6}= -2 \left(38 m_{(1)}^2 m_{(2)}-15 m_{(2)}^3\right)\nonumber\, ,\\
&g_{L7}= \frac{3}{10} \left(122 m_{(1)}^2 m_{(2)}-75 m_{(2)}^3\right)\, , &&g_{L8}= \frac{3}{10} \left(10 m_{(1)}^2 m_{(2)}-77 m_{(2)}^3\right)\nonumber\, ,\\
&g_{L9}= \frac{1}{10} (-3) \left(77 m_{(1)}^3-10 m_{(1)} m_{(2)}^2\right)\, , &&g_{L10}= \frac{1}{10} (-3) \left(75 m_{(1)}^3-122 m_{(1)} m_{(2)}^2\right)\nonumber\, ,\\
&g_{L11}= 2 \left(15 m_{(1)}^3-38 m_{(1)} m_{(2)}^2\right)\, , &&g_{L12}= \frac{122}{5} m_{(1)} m_{(2)}^2\, .
\end{align}
Following these steps, we ultimately obtain the full effective Hamiltonian free of divergences and logarithms for the non-spinning sector till 3PN and for the spin-orbit sector till \NNNLO. 

\section{Results}\label{sec_results}

To begin, we first define a set of dimensionless variables to express the Hamiltonian 
in a compact form. 
We introduce the total mass $M=m_{(1)}+m_{(2)}$, the reduced mass of the two body system $\mu=m_{(1)}m_{(2)}/M$, the mass ratio $q=m_{(1)}/m_{(2)}$, and the symmetric mass ratio $\nu=\mu/M$. These are related as follows
\begin{align}
\nu=\frac{m_{(1)} m_{(2)}}{M^2}=\frac{\mu}{M}=\frac{q}{(1+q)^2}\, .
\end{align}
Additionally, we express the results in the center of mass (COM) frame of reference and define the momentum in COM frame as $\textbf{p} \equiv \textbf{p}_{(1)}=-\textbf{p}_{(2)}$.
In the COM, the orbital angular momentum is defined as $\textbf{L}=(\textbf{r}\times \textbf{p})$.
So, we can write $p^2 = p_r^2 + L^2 / r^2$, where $p_r=\textbf{p}\cdot \textbf{n}$, $p\equiv |\textbf{p}|$ and $L\equiv |\textbf{L}|$.
All the variables are rescaled to write the expressions in the form of dimensionless 
parameters
\begin{align}
\widetilde{\textbf{p}}=\frac{\textbf{p}}{\mu c} \quad\, ,\quad\quad \widetilde{\textbf{r}}=\frac{\textbf{r}~c^2}{G_N M} \quad\, ,\quad\quad \widetilde{\textbf{L}}=\frac{\textbf{L}~c}{G_N M \mu} \quad\, ,\quad\quad \widetilde{\textbf{S}}_{(a)}=\frac{\textbf{S}_{(a)}}{G_N M \mu} \quad\, ,\quad\quad \widetilde{\mathcal{H}}=\frac{\mathcal{H}}{\mu c^2}\, .
\end{align}

The total EFT Hamiltonian can be written as 
\begin{align}
\widetilde{\mathcal{H}} = \widetilde{\mathcal{H}}_{\rm pp} + \widetilde{\mathcal{H}}_{\rm SO}\ ,
\end{align}
where, 
\begin{align}\label{eq_Ham_pp}
\widetilde{\mathcal{H}}_{\rm pp} &= \widetilde{\mathcal{H}}_{\text{0PN}}+\left(\frac{1}{c^2}\right)\widetilde{\mathcal{H}}_{\text{1PN}}+\left(\frac{1}{c^4}\right)\widetilde{\mathcal{H}}_{\text{2PN}}+\left(\frac{1}{c^6}\right)\widetilde{\mathcal{H}}_{\text{3PN}}+\mathcal{O}\left(\frac{1}{c^8}\right) \ , \\
\label{eq_Ham_SO}
\widetilde{\mathcal{H}}_{\rm SO} &= \left(\frac{1}{c^3}\right)\widetilde{\mathcal{H}}_{\text{LO}}+\left(\frac{1}{c^5}\right)\widetilde{\mathcal{H}}_{\text{NLO}}+\left(\frac{1}{c^7}\right)\widetilde{\mathcal{H}}_{\rm N^2LO}+\left(\frac{1}{c^9}\right)\widetilde{\mathcal{H}}_{\rm N^3LO}+\mathcal{O}\left(\frac{1}{c^{11}}\right)\, .
\end{align}
In the non-spinning part, the Hamiltonian till 3PN is known in the literature~\cite{Foffa:2011ub} 
and in the spinning sector the Hamiltonian is known till NNLO~\cite{Levi:2015uxa}. 
The novel result of the $\tilde{\mathcal{H}}_{\rm N^3LO}$ in the EFT approach in the COM frame is presented here 
\begin{align}
\widetilde{\mathcal{H}}_{\rm N^3LO}&= 
\frac{\left(\widetilde{\textbf{S}}_{(1)}\cdot \widetilde{\textbf{L}}\right)}{q} \Bigg\{
\frac{\left(10864+2025 \pi ^2\right) \nu ^3}{800 \widetilde{r}^6}+\frac{\left(17702+2895 \pi ^2\right) \nu ^2}{960 \widetilde{r}^6}-\frac{181 \nu }{8 \widetilde{r}^6}\nonumber\\
&\quad\quad\quad\quad\quad+\widetilde{p}_r^2\left(\frac{\left(53153+4050 \pi ^2\right) \nu ^3}{400 \widetilde{r}^5}+\frac{\left(60989+225 \pi ^2\right) \nu ^2}{480 r^5}-\frac{10 \nu }{\widetilde{r}^5}\right) \nonumber\\
&\quad\quad\quad\quad\quad+\widetilde{p}_r^4\left(-\frac{263 \nu ^4}{32 \widetilde{r}^4}-\frac{29 \nu ^3}{\widetilde{r}^4}+\frac{741 \nu ^2}{32 \widetilde{r}^4}-\frac{13 \nu }{4 \widetilde{r}^4}\right) 
+\widetilde{p}_r^6\left(\frac{425 \nu ^4}{32 \widetilde{r}^3}-\frac{745 \nu ^3}{64 \widetilde{r}^3}+\frac{223 \nu ^2}{64 \widetilde{r}^3}-\frac{45 \nu }{128 \widetilde{r}^3}\right) \nonumber\\
&\quad\quad\quad\quad\quad+\widetilde{L}^2\bigg(
-\frac{3 \left(-412+675 \pi ^2\right) \nu ^3}{800 \widetilde{r}^7}+\frac{\left(59116-225 \pi ^2\right) \nu ^2}{1920 \widetilde{r}^7}-\frac{85 \nu }{8 \widetilde{r}^7}\nonumber\\
&\quad\quad\quad\quad\quad\quad\quad+\widetilde{p}_r^2\left(-\frac{17 \nu ^4}{4 \widetilde{r}^6}-\frac{4257 \nu ^3}{64 \widetilde{r}^6}+\frac{2253 \nu ^2}{32 \widetilde{r}^6}-\frac{13 \nu }{2 \widetilde{r}^6}\right) \nonumber\\
&\quad\quad\quad\quad\quad\quad\quad+\widetilde{p}_r^4\left(\frac{339 \nu ^4}{32 \widetilde{r}^5}-\frac{1479 \nu ^3}{64 \widetilde{r}^5}+\frac{597 \nu ^2}{64 \widetilde{r}^5}-\frac{135 \nu }{128 \widetilde{r}^5}\right) 
\bigg) \nonumber\\
&\quad\quad\quad\quad\quad+\widetilde{L}^4\bigg(-\frac{29 \nu ^4}{32 \widetilde{r}^8}-\frac{789 \nu ^3}{64 \widetilde{r}^8}+\frac{163 \nu ^2}{16 \widetilde{r}^8}-\frac{13 \nu }{4 \widetilde{r}^8}
+\widetilde{p}_r^2\left(\frac{153 \nu ^4}{32 \widetilde{r}^7}-\frac{1053 \nu ^3}{64 \widetilde{r}^7}+\frac{525 \nu ^2}{64 \widetilde{r}^7}-\frac{135 \nu }{128 \widetilde{r}^7}\right) \bigg) \nonumber\\
&\quad\quad\quad\quad\quad+\widetilde{L}^6\left(\frac{29 \nu ^4}{32 \widetilde{r}^9}-\frac{249 \nu ^3}{64 \widetilde{r}^9}+\frac{151 \nu ^2}{64 \widetilde{r}^9}-\frac{45 \nu }{128 \widetilde{r}^9}\right) 
\Bigg\}\nonumber\\
&+\left(\widetilde{\textbf{S}}_{(1)}\cdot \widetilde{\textbf{L}}\right) \Bigg\{
\frac{\left(10864+2025 \pi ^2\right) \nu ^3}{800 \widetilde{r}^6}+\frac{\left(-85957+8400 \pi ^2\right) \nu ^2}{3600 \widetilde{r}^6}-\frac{34 \nu }{\widetilde{r}^6}\nonumber\\
&\quad\quad\quad\quad\quad+\widetilde{p}_r^2\left(\frac{81 \left(1451+100 \pi ^2\right) \nu ^3}{800 \widetilde{r}^5}-\frac{\left(52076+5175 \pi ^2\right) \nu ^2}{1200 \widetilde{r}^5}-\frac{37 \nu }{\widetilde{r}^5}\right) \nonumber\\
&\quad\quad\quad\quad\quad+\widetilde{p}_r^4\left(-\frac{263 \nu ^4}{32 \widetilde{r}^4}-\frac{599 \nu ^3}{16 \widetilde{r}^4}+\frac{561 \nu ^2}{32 \widetilde{r}^4}-\frac{3 \nu }{\widetilde{r}^4}\right) 
+\widetilde{p}_r^6\left(\frac{2025 \nu ^4}{128 \widetilde{r}^3}-\frac{325 \nu ^3}{32 \widetilde{r}^3}+\frac{153 \nu ^2}{128 \widetilde{r}^3}\right) \nonumber\\
&\quad\quad\quad\quad\quad+\widetilde{L}^2\bigg(
+\frac{\left(5461-2025 \pi ^2\right) \nu ^3}{800 \widetilde{r}^7}+\frac{\left(490976+5175 \pi ^2\right) \nu ^2}{4800 \widetilde{r}^7}+\frac{11 \nu }{\widetilde{r}^7}\nonumber\\
&\quad\quad\quad\quad\quad\quad\quad+\widetilde{p}_r^2\left(-\frac{17 \nu ^4}{4 \widetilde{r}^6}-\frac{2677 \nu ^3}{32 \widetilde{r}^6}+\frac{2115 \nu ^2}{32 \widetilde{r}^6}-\frac{2 \nu }{\widetilde{r}^6}\right) +\widetilde{p}_r^4\left(\frac{1671 \nu ^4}{128 \widetilde{r}^5}-\frac{471 \nu ^3}{32 \widetilde{r}^5}+\frac{375 \nu ^2}{128 \widetilde{r}^5}\right) 
\bigg) \nonumber\\
&\quad\quad\quad\quad\quad+\widetilde{L}^4\left(-\frac{29 \nu ^4}{32 \widetilde{r}^8}-\frac{327 \nu ^3}{16 \widetilde{r}^8}-\frac{71 \nu ^2}{32 \widetilde{r}^8}+\frac{\nu }{\widetilde{r}^8}+\widetilde{p}_r^2\left(\frac{777 \nu ^4}{128 \widetilde{r}^7}-\frac{387 \nu ^3}{32 \widetilde{r}^7}+\frac{471 \nu ^2}{128 \widetilde{r}^7}\right) \right) \nonumber\\
&\quad\quad\quad\quad\quad+ \widetilde{L}^6\left(\frac{151 \nu ^4}{128 \widetilde{r}^9}-\frac{101 \nu ^3}{32 \widetilde{r}^9}+\frac{109 \nu ^2}{128 \widetilde{r}^9}\right)
\Bigg\}\nonumber\\
& + (1\leftrightarrow 2)\, .
\end{align}
Here, $(1\leftrightarrow 2)$ implies changing the label $1$ and $2$ on the spin variables along with $q\leftrightarrow 1/q$.

The computation of the previously known Hamiltonians, 
together with the novel computation of the $\tilde{\mathcal{H}}_{\rm N^3LO}$,
allows us to obtain
%for the first time,
the complete expression for the spin-orbit \NNNLO Hamiltonian within the EFT framework.
We provide the known Hamiltonians in appendix \ref{app_Ham}.
The Hamiltonian in the generic frame is also provided in the ancillary file \texttt{Hamiltonian.m} with this article.

\section{Computation of observables with spin}
\label{sec_compputing_observables}

The obtained Hamiltonian is still gauge dependent 
as it depends on the radial coordinate, 
we compute gauge invariant observable
to compare with the literature~\cite{Antonelli:2020aeb}.
In this section, we describe the procedure for the computation of two gauge invariant observable: the binding energy for circular orbit with aligned spin configuration and the scattering angle with aligned spin configuration. 

For this, we work in the COM defined by $\textbf{p}_{(1)}+\textbf{p}_{(2)}=0$ as given in section \ref{sec_results}. 
With this, we assume that the spins are aligned to the direction of the orbital angular momentum of the binary system. 
Such aligned spin configuration is given by
\begin{align}
\textbf{S}_{(a)}\cdot \textbf{r} = \textbf{S}_{(a)}\cdot \textbf{p} = 0 \implies \textbf{S}_{(a)}\cdot( \textbf{r}\times \textbf{p}) = S_{(a)} L\, ,
\end{align}
where, $L=|\textbf{L}|$ and $S_{(a)}=|\textbf{S}_{(a)}|$.

\subsection{Binding energy for circular orbits with aligned spins}
We compute the gauge invariant relation between the binding energy and the orbital frequency for circular orbits by removing the dependence on the radial coordinate.
For circular orbits, we know
\begin{align}
p_r=0 %\implies p^2&=\frac{L^2}{r^2} \\
\quad\quad\quad\text{and} \quad\quad\quad
\frac{dp_r}{dt} = 0\, ,
\end{align}
%Using the above, we derive
which implies
\begin{equation}
    \frac{\partial \widetilde{\mathcal{H}}(\widetilde{r},\widetilde{L},\widetilde{S}_{(a)})}{\partial \widetilde{r}}=0\, .
\end{equation}
We invert the above relation to express $\widetilde{r}$ as a function of $\widetilde{L}$ %Using this we can express the Hamiltonian as $\widetilde{H}(\widetilde{L},\widetilde{S}_{(a)})$.
and we define the orbital frequency as
\begin{align}
\widetilde{\omega}=\frac{\partial \widetilde{\mathcal{H}}(\widetilde{L},\widetilde{S}_{(a)})}{\partial \widetilde{L}}\, .
\end{align}
We invert the above relation again to express $\widetilde{L}$ as a function of $\widetilde{\omega}$.
Additionally, we
define a gauge invariant PN parameter $x=\widetilde{\omega}^{2/3}$ .
Using the above two equations, we express $\tilde{L}$ as a function of $x$, and substituting this in the Hamiltonian, we obtain the gauge invariant binding energy of the circular orbits in the aligned spin configuration $\tilde{E}(x,\tilde{S}_{(a)})$.
Following the above procedure with the Hamiltonian given in section \ref{sec_results} we obtain
\begin{align}\label{eq_BE}
E(x,\widetilde{S}_{(a)})= E_{\text{pp}}(x)+E_{\text{SO}}(x,\widetilde{S}_{(a)})\, ,
\end{align} 
where,
\begin{align}
E_{\text{pp}}(x)&= -x \frac{1}{2} + x^2 \left\{\frac{3}{8}+\frac{\nu }{24}\right\} + x^3 \left\{\frac{27}{16}-\frac{19  }{16}\nu+\frac{1}{48}\nu ^2\right\} \nonumber\\
&+ x^4 \left\{\frac{675}{128}+\left(-\frac{34445}{1152}+\frac{205 \pi ^2}{192}\right)\nu+\frac{155 }{192}\nu ^2+\frac{35 }{10368}\nu ^3\right\}\, ,
\end{align}
and
\begin{align}
\label{eq:BE_SO_N3LO}
E_{\text{SO}}(x,\widetilde{S})&= x^{5/2} \Bigg\{
S^*  \left(-\nu\right) 
+ S  \left(-\frac{4}{3} \nu\right) 
\Bigg\} \nonumber\\
&+x^{7/2} \Bigg\{
S^* \left(-\frac{3}{2}\nu+\frac{5}{3}\nu ^2\right)  
+S \left(-4 \nu +\frac{31}{18} \nu ^2\right)  
\Bigg\} \nonumber\\
&+x^{9/2} \Bigg\{
S^*\left(-\frac{27  }{8}\nu+\frac{39 }{2}\nu ^2-\frac{5 }{8}\nu ^3\right) 
+S \left(-\frac{27  }{2}\nu+\frac{211 }{8}\nu ^2-\frac{7 }{12}\nu ^3\right) 
\Bigg\} \nonumber\\
&+x^{11/2} \Bigg\{
S^* \left(-\frac{135  }{16}\nu+\frac{565 }{8}\nu ^2-\frac{1109 }{24}\nu ^3-\frac{25 }{324}\nu ^4\right) \nonumber\\
&\quad\quad\quad +S \left(-45 \nu +\left(\frac{19679}{144}+\frac{29 \pi ^2}{24}\right) \nu ^2 -\frac{1979 }{36}\nu ^3-\frac{265 }{3888}\nu ^4\right) 
\Bigg\} \, ,
\end{align}
 $S=\widetilde{S}_{(1)}+\widetilde{S}_{(2)}$, and $S^\star=\widetilde{S}_{(1)}/q+\widetilde{S}_{(2)}q$.
Our result given in equation \eqref{eq:BE_SO_N3LO} agrees with the result 
given in equation (10) of \cite{Antonelli:2020aeb}.

\subsection{Scattering angle with aligned spins}
In this section, we study the scattering angle considering an aligned spin binary system following \cite{Vines:2018gqi}. 
First, we re-scale the spin variables as $a_{(a)}=S_{(a)}/m_{(a)}$.
In the COM, the Hamiltonian $\mathcal{H}$ is expressed as a function of $p_r$, $L$, $r$, and $S{(a)}$ and inverting that we obtain
\begin{align}
p_r=p_r(\mathcal{H},L,r,S_{(a)})\, .
\end{align}
Then the scattering angle $\chi$ is given by 
\begin{align}\label{eq_scattering_angle_1}
\chi(\mathcal{H},L,S_{(a)})=-\int dr \frac{\partial p_r(\mathcal{H},L,r,S_{(a)})}{\partial L}  - \pi\, .
\end{align}
Now, the total center of mass energy ($E$) and the total energy per 
total rest mass ($\Gamma$) is given by 
\begin{align}\label{eq_H_for_gamma}
\mathcal{H}=E&=Mc^2\sqrt{1+2\nu(\gamma-1)}\, ,\\
\Gamma&=\frac{\mathcal{H}}{M c^2}=\sqrt{1+2\nu(\gamma-1)}\, ,
\end{align}
where $\gamma$ is the Lorentz factor.
We invert the above relation to express the Lorentz factor $\gamma$ in terms of $\Gamma$, and we obtain
\begin{align}\label{eq_gamma_for_v}
\gamma=\frac{1}{\sqrt{1-v^2/c^2}}=1+\frac{\Gamma^2-1}{2\nu}\, ,
\end{align}
where, $v\equiv|\dot{\textbf{r}}|$ is the relative velocity of the compact objects.
%as measured from infinity.
Moreover, the total angular momentum $L$ can be expressed in terms of the impact parameter $b$ and the aligned spin configurations $a_{(+)}$ and $a_{(-)}$ as 
\begin{align}\label{eq_L_for_b}
L=\frac{\mu\gamma v b}{\Gamma} + Mc \left(\frac{\Gamma-1}{2}\right) \left(a_+ - \frac{\delta}{\Gamma} a_-\right)\, ,
\end{align}
where, $\delta=(m_{(1)}-m_{(2)})/M$, $a_{(+)}=a_{(1)}+a_{(2)}$ and $a_{(-)}=a_{(1)}-a_{(2)}$. 
We use equation \eqref{eq_H_for_gamma} and \eqref{eq_gamma_for_v} to trade $H$ for $v$
and using equation \eqref{eq_L_for_b} we trade $L$ for $b$.
This allows us to express the scattering angle as
\begin{align}
\chi(v,b,S_{(a)})=-\frac{\gamma}{\mu\gamma v}\int dr \frac{\partial p_r(v,b,r,S_{(a)})}{\partial b}  - \pi\, .
\end{align}
Now, applying the above procedure with the Hamiltonian given in section \ref{sec_results}, we obtain the scattering angle computed in the COM for aligned spins, which can be expressed as
\begin{align}\label{eq_SA}
\chi(v,b,S_{(a)})= \chi_{\text{pp}}(v,b)+\chi_{\text{SO}}(v,b,S_{(a)})\, ,
\end{align} 
where,
\begin{align}
 \frac{\chi_{\text{pp}}}{\Gamma}&=  
\left(\frac{G_NM}{v^2b}\right)\left\{2+2\left(\frac{v^2}{c^2}\right) +\mathcal{O}\left(\frac{v^8}{c^8}\right)\right\}\nonumber\\
&+\pi \left(\frac{G_NM}{v^2b}\right)^2 \left\{3\left(\frac{v^2}{c^2}\right)+\frac{3}{4}\left(\frac{v^4}{c^4}\right) +\mathcal{O}\left(\frac{v^8}{c^8}\right)\right\}\nonumber\\
&+ \left(\frac{G_NM}{v^2b}\right)^3 \left\{-\frac{2}{3}+2\frac{15-\nu}{3}\left(\frac{v^2}{c^2}\right)+\frac{60-13\nu}{2}\left(\frac{v^4}{c^4}\right)+\frac{40-277\nu}{12}\left(\frac{v^6}{c^6}\right) +\mathcal{O}\left(\frac{v^8}{c^8}\right)\right\}\nonumber\\
&+\pi\left(\frac{G_NM}{v^2b}\right)^4 \left\{15\frac{7-2\nu}{4}\left(\frac{v^4}{c^4}\right)+\left(\frac{105}{4}-\frac{437}{8}\nu+\frac{123}{128}\pi^2\nu\right)\left(\frac{v^6}{c^6}\right) +\mathcal{O}\left(\frac{v^8}{c^8}\right)\right\}\nonumber\\
&+\mathcal{O}(G_N^5)\, ,
\end{align}
and
\begin{align}
\label{eq:SA_SO_N3LO}
\frac{\chi_{\text{SO}}}{\Gamma}=  \frac{v}{bc} 
\begin{bmatrix}%
a_{(+)}  & ~\delta a_{(-)}
\end{bmatrix}
\cdot&\Bigg(
\left(\frac{G_NM}{v^2b}\right)\left\{  \begin{bmatrix}-4 \\ 0\end{bmatrix} +\mathcal{O}\left(\frac{v^8}{c^8}\right)\right\}\nonumber\\
&+\pi \left(\frac{G_NM}{v^2b}\right)^2 \left\{ -\frac{1}{2}  \begin{bmatrix}7 \\ 1\end{bmatrix}-\frac{3}{4}  \begin{bmatrix}7 \\ 1\end{bmatrix} \left(\frac{v^2}{c^2}\right) +\mathcal{O}\left(\frac{v^8}{c^8}\right)\right\}\nonumber\\
&+ \left(\frac{G_NM}{v^2b}\right)^3 \bigg\{ -2  \begin{bmatrix}5 \\ 1\end{bmatrix}-20  \begin{bmatrix}5-\nu/2 \\ 1\end{bmatrix} \left(\frac{v^2}{c^2}\right) -10  \begin{bmatrix}5-77\nu/20 \\ 1\end{bmatrix} \left(\frac{v^4}{c^4}\right)\nonumber\\
&\quad\quad\quad\quad\quad\quad\quad+ \frac{1}{4} \begin{bmatrix}177\nu \\ 0\end{bmatrix} \left(\frac{v^6}{c^6}\right)+\mathcal{O}\left(\frac{v^8}{c^8}\right)\bigg\}\nonumber\\
&+ \pi\left(\frac{G_NM}{v^2b}\right)^4 \bigg\{ \frac{3}{4}  \begin{bmatrix}-91+13\nu \\ -21+\nu\end{bmatrix}\left(\frac{v^2}{c^2}\right) -\frac{1}{8}  \begin{bmatrix}1365-777\nu \\ 315-45\nu\end{bmatrix} \left(\frac{v^4}{c^4}\right) \nonumber\\
&\quad\quad\quad\quad\quad\quad\quad-\frac{1}{32}  \begin{bmatrix}1365-\left(\frac{23717}{3}-\frac{733}{8}\pi^2\right)\nu \\ 315-\left(\frac{257}{3}+\frac{251}{8}\pi^2\right)\nu \end{bmatrix} \left(\frac{v^6}{c^6}\right)+\mathcal{O}\left(\frac{v^8}{c^8}\right)\bigg\}\Bigg)\nonumber\\
&+\mathcal{O}(G_N^5)\, .
\end{align}
Our result given in equation \eqref{eq:SA_SO_N3LO} agrees with the result given in equation (7) of \cite{Antonelli:2020aeb}.
We note that the \NNNLO spin-orbit scattering angle contains all gauge-invariant information of the corresponding Lagrangian or Hamiltonian~\cite{Antonelli:2020ybz}, albeit restricted to the center-of-mass frame.
%Hence for a complete check of our Lagrangian or Hamiltonian only a verification of invariance under Lorentz boosts is missing, which we leave for future work.

\section{Conclusion}\label{sec_Conclusion}
In this work, we presented the complete evaluation of \NNNLO Post-Newtonian (PN) correction to the the spin-orbit Hamiltonian for rapidly rotating compact objects, within the effective field theory diagrammatic approach of General Relativity. 

The required Feynman diagrams in momentum space were automatically generated 
using EFT Feynman rules for the interaction of spinning compact objects. They were algebraically decomposed in terms of master integrals, by means of integration by parts identities for dimensionally regulated integrals, whose expressions were available in the literature. 
The contribution of each diagram to the Lagrangian, namely to the effective potential, was found after taking the Fourier transform to position space, and series expanding around $d=3+\epsilon$ space dimension. 
The Hamiltonian is finally found by means of Legendre transform, followed by suitable canonical transformations, required for eliminating residual non-physical divergences and spurious logarithmic behaviours.

We computed the gauge invariant relation between binding energy and the orbital frequency in the circular orbit with aligned spins as well as the 
the scattering angle with aligned spins and they were found in agreement with the results available in the literature, previously obtained using the self-force formalism.

Our results complete the description of coalescing rapidly rotating binary systems up to 4.5 PN order within the spin-orbit sector.

The techniques and the in-house code developed for the current project are very flexible, and can be applied to other interesting problems, at higher order in the PN expansion, for both non-spinning and spinning compact objects.

\paragraph{Note added} During the completion of this work, a similar study has appeared in \cite{Kim:2022pou}.

\subsection*{Acknowledgements}
We would like to thank Jonathan Ronca and William J. Torres Bobadilla for partial checks on the non-spinning sector and interesting discussions regarding the reduction methods for Feynman integrals.
We would like to thank Jan Plefka for suggesting useful corrections in the earlier version of the paper.
The work of M.K.M is supported by Fellini - Fellowship for Innovation at INFN funded by the European Union's Horizon 2020 research and innovation programme under the Marie Sk{\l}odowska-Curie grant agreement No 754496. 
RP is grateful to IISER Bhopal for the fellowship. RP's research is funded by the Deutsche Forschungsgemeinschaft (DFG, German Research Foundation), Projektnummer 417533893/GRK2575 “Rethinking Quantum Field Theory”.  
\appendix
%\addtocontents{toc}{\protect\setcounter{tocdepth}{1}}
%\newpage
\section{Notation and convention} \label{app_notation_and_convention}

\begin{subequations}   
	\begin{eqnarray} 
	\textrm{Spacetime metric}\quad \quad && \eta_{\mu \nu }=(1,-1,-1,-1)\\
	\textrm{4 dimensional indices}\quad \quad && \mu,\nu \\
	\textrm{3 dimensional indices}\quad \quad && i,j 
	\\
	\textrm{Compact object label}\quad \quad&& _{(a)} \quad \textrm{where } a=\{1,2\}\\
	\textrm{Time derivative}\quad \quad&& \dot{}\\
	\textrm{Position of $a^{\text{th}}$ object}\quad \quad&&\textbf{x}_{(a)}\\
	\textrm{Velocity of $a^{\text{th}}$ object}\quad \quad&&\textbf{v}_{(a)}\equiv\dot{\textbf{x}}_{(a)}\\
	\textrm{Acceleration of $a^{\text{th}}$ object}\quad \quad&&\textbf{a}_{(a)}\equiv\ddot{\textbf{x}}_{(a)}\\
	\textrm{Separation vector for binary}\quad \quad&&\textbf{r}\equiv\textbf{x}_{(a)}-\textbf{x}_{(a)}\\
	\textrm{Separation distance for binary}\quad \quad&&r\equiv|\textbf{r}|\\
	\textrm{Separation unit vector for binary}\quad \quad&&\textbf{n}\equiv\frac{\textbf{r}}{r}\\
	\textrm{Angular momentum of the binary}\quad \quad&& \textbf{L}\equiv(\textbf{r}\times \textbf{p})\\
    \textrm{Spin vector of $a^{\text{th}}$ object}\quad \quad&& \textbf{S}_{(a)}^i\equiv\bm{\epsilon}^{ijk} \textbf{S}_{(a)}^{jk}\\
	\int_p\quad \quad&& \int \frac{d^dp}{(2\pi)^d}\\
	\int_\textbf{p}\quad \quad&& \int \frac{d^3p}{(2\pi)^3}\\
	\textrm{Center of mass coordinates}\quad \quad&&\textbf{p}_{(1)}+\textbf{p}_{(2)}=0\\
	\textrm{Circular orbits}\quad \quad&&p_r\equiv\textbf{p}\cdot \textbf{n}=0 \quad\quad\textrm{and}\quad\quad\dot{p}_r=0\\
	\textrm{Aligned spins}\quad \quad&&\textbf{S}_{(a)}\cdot \textbf{r} =\textbf{S}_{(a)}\cdot \textbf{p} = 0
	\end{eqnarray} 
\end{subequations}

\section{Relevant Integrals}
\label{app_master_and_fourier_ints}

In this appendix, we present all the integrals used in the computation of results given in section \ref{sec_results}. We first give the master integrals at different loops that are used in the multi-loop methods. Then we give the Fourier integral used to obtain the effective potential as described computational algorithm described in figure \ref{fig_flowchart_of_routine}.

\subsection{Master Integrals}\label{sec_masterints}

\begin{figure}[H]
	\centering
	\begin{subfigure}{0.19\textwidth}
		\centering
		\begin{tikzpicture}[line width=1 pt, scale=0.5]
		\draw (-2,0)--(-1,0);
		\draw (2,0)--(1,0);
		\draw (0,0) circle (1);
		\end{tikzpicture}
		\caption{$M_{1,1}$}
		%		\label{}
	\end{subfigure}
	\caption{One loop master integrals}
	\label{}
\end{figure}
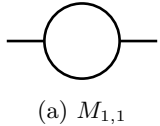
\begin{align}
M_{1,1}&= \int_{k_1} \frac{1}{k_1^2}\frac{1}{(k_1-p)^2}\\
%&=2^{-d} \pi ^{-\frac{d}{2}} p^{d-4} \frac{\Gamma \left(2-\frac{d}{2}\right) \Gamma \left(\frac{d}{2}-1\right)^2}{\Gamma (d-2)}
&= (4 \pi)^{-\frac{d}{2}} p^{d-4} \frac{\Gamma \left(2-\frac{d}{2}\right) \Gamma \left(\frac{d}{2}-1\right)^2}{\Gamma (d-2)}
\end{align}

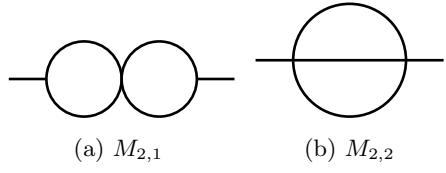
\begin{figure}[H]
	\centering
	\begin{subfigure}{0.19\textwidth}
		\centering
		\begin{tikzpicture}[line width=1 pt, scale=0.5]
		\draw (-2,0)--(-1,0);
		\draw (4,0)--(3,0);
		\draw (0,0) circle (1);
		\draw (2,0) circle (1);
		\end{tikzpicture}
		\caption{$M_{2,1}$}
		%		\label{}
	\end{subfigure}
	\begin{subfigure}{0.19\textwidth}
		\centering
		\begin{tikzpicture}[line width=1 pt, scale=0.5]
		\draw (-2.5,0)--(-1.5,0);
		\draw (2.5,0)--(1.5,0);
		\draw (-1.5,0)--(1.5,0);
		\draw (0,0) circle (1.5);
		\end{tikzpicture}	
		\caption{$M_{2,2}$}
		%		\label{}
	\end{subfigure}
	\caption{Two loop master integrals}
	\label{}
\end{figure}
\begin{align}
M_{2,1}&= \int_{k_1,k_2} \frac{1}{k_1^2}\frac{1}{(k_1-p)^2}\frac{1}{(k_1+k_2)^2}\frac{1}{(k_1+k_2-p)^2}\\
%&=2^{2-2 d} \pi ^{-d} p^{2 d-8} \frac{1}{(d-4)^2 (d-3)^2 }\frac{\Gamma \left(3-\frac{d}{2}\right)^2 \Gamma \left(\frac{d}{2}-1\right)^4}{\Gamma (d-3)^2}
&=4\, (4 \pi)^{-d} p^{2 d-8} \frac{1}{(d-4)^2 (d-3)^2 }\frac{\Gamma \left(3-\frac{d}{2}\right)^2 \Gamma \left(\frac{d}{2}-1\right)^4}{\Gamma (d-3)^2}
\end{align}
\begin{align}
M_{2,2}&= \int_{k_1,k_2} \frac{1}{k_1^2}\frac{1}{k_2^2}\frac{1}{(k_1+k_2-p)^2}\\
%&=2^{2-2 d} \pi ^{-d} p^{2 d-6} \frac{1}{(d-4) (d-3) (3 d-10) (3 d-8) } \frac{\Gamma(5-d) \Gamma \left(\frac{d}{2}-1\right)^3}{\Gamma \left(\frac{3 d}{2}-5\right)}
&=4 \, (4 \pi)^{-d} p^{2 d-6} \frac{1}{(d-4) (d-3) (3 d-10) (3 d-8) } \frac{\Gamma(5-d) \Gamma \left(\frac{d}{2}-1\right)^3}{\Gamma \left(\frac{3 d}{2}-5\right)}
\end{align}

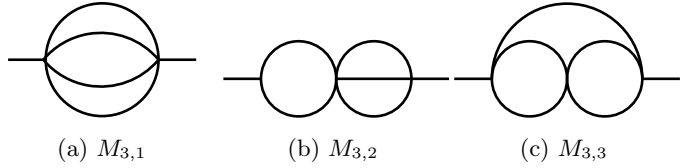
\begin{figure}[H]
	\centering
	\begin{subfigure}{0.19\textwidth}
		\centering
		\begin{tikzpicture}[line width=1 pt, scale=0.5]
		\draw (-2.5,0)--(-1.5,0);
		\draw (2.5,0)--(1.5,0);
		\draw (1.5,0) arc (40:140:2);
		\draw (1.5,0) arc (-40:-140:2);
		\draw (0,0) circle (1.5);
		\end{tikzpicture}
		\caption{$M_{3,1}$}
		%		\label{}
	\end{subfigure}
	\begin{subfigure}{0.19\textwidth}
		\centering
		\begin{tikzpicture}[line width=1 pt, scale=0.5]
		\draw (-2,0)--(-1,0);
		\draw (4,0)--(3,0);
		\draw (0,0) circle (1);
		\draw (2,0) circle (1);
		\draw (1,0)--(3,0);
		\end{tikzpicture}
		\caption{$M_{3,2}$}
		%		\label{}
	\end{subfigure}
	\begin{subfigure}{0.19\textwidth}
		\centering
		\begin{tikzpicture}[line width=1 pt, scale=0.5]
		\draw (-2,0)--(-1,0);
		\draw (4,0)--(3,0);
		\draw (0,0) circle (1);
		\draw (2,0) circle (1);
		\draw (3,0) arc (0:180:2);
		\end{tikzpicture}		
		\caption{$M_{3,3}$}
		%		\label{}
	\end{subfigure}
	\caption{Three loop master integrals}
	\label{}
\end{figure}
\begin{align}
M_{3,1}&= \int_{k_1,k_2,k_3} \frac{1}{k_1^2}\frac{1}{k_2^2}\frac{1}{k_3^2}\frac{1}{(k_1+k_2+k_3-p)^2}\\
%&=2^{-3 d} \pi ^{-\frac{3 d}{2}} p^{3 d-8}\frac{ \Gamma \left(4-\frac{3 d}{2}\right) \Gamma \left(\frac{d}{2}-1\right)^4}{\Gamma (2 d-4)}
&= (4 \pi)^{-\frac{3 d}{2}} p^{3 d-8}\frac{ \Gamma \left(4-\frac{3 d}{2}\right) \Gamma \left(\frac{d}{2}-1\right)^4}{\Gamma (2 d-4)}
\end{align}
\begin{align}
M_{3,2}&= \int_{k_1,k_2,k_3} \frac{1}{k_1^2}\frac{1}{k_2^2}\frac{1}{(k_1-k_3)^2}\frac{1}{(k_1+k_2-p)^2}\frac{1}{(k_1-k_3-p)^2}\\
%&=2^{-3 d} \pi ^{-\frac{3 d}{2}} p^{3 d-10}\frac{ \Gamma (3-d) \Gamma \left(2-\frac{d}{2}\right) \Gamma \left(\frac{d}{2}-1\right)^5}{\Gamma (d-2) \Gamma \left(\frac{3 d}{2}-3\right)}
&=(4 \pi)^{-\frac{3 d}{2}} p^{3 d-10}\frac{ \Gamma (3-d) \Gamma \left(2-\frac{d}{2}\right) \Gamma \left(\frac{d}{2}-1\right)^5}{\Gamma (d-2) \Gamma \left(\frac{3 d}{2}-3\right)}
\end{align}
\begin{align}
M_{3,3}&= \int_{k_1,k_2,k_3} \frac{1}{k_1^2}\frac{1}{k_2^2}\frac{1}{k_3^2}\frac{1}{(k_1+k_2+k_3)^2}\frac{1}{(k_1+k_3-p)^2}\\
%&=2^{-3 d} \pi ^{-\frac{3 d}{2}} p^{3 d-10}\frac{ \Gamma \left(5-\frac{3 d}{2}\right) \Gamma \left(2-\frac{d}{2}\right)^2 \Gamma \left(\frac{d}{2}-1\right)^5 \Gamma \left( \frac{3 d}{2}-4\right)}{\Gamma (4-d) \Gamma (d-2)^2 \Gamma (2 d-5)}
&=(4 \pi)^{-\frac{3 d}{2}} p^{3 d-10}\frac{ \Gamma \left(5-\frac{3 d}{2}\right) \Gamma \left(2-\frac{d}{2}\right)^2 \Gamma \left(\frac{d}{2}-1\right)^5 \Gamma \left( \frac{3 d}{2}-4\right)}{\Gamma (4-d) \Gamma (d-2)^2 \Gamma (2 d-5)}
\end{align}

\subsection{Fourier Integrals}\label{sec_fourierints}
The expression for the scalar Fourier integral is given by 
\begin{align}
\int_k  e^{\iimg \textbf{k}^i  \textbf{x}_i} \left( \textbf{k}^i \textbf{k}_i \right)^{-\alpha} = \frac{2^{-2 \alpha } \pi ^{-\frac{d}{2}} \Gamma \left(\frac{d}{2}-\alpha \right) \left(\textbf{x}^i \textbf{x}_i \right)^{\alpha -\frac{d}{2}}}{\Gamma (\alpha )}\, .
\end{align}
which is for generic dimensions but only in Euclidean signature. The required tensor Fourier integrals are then obtained by taking derivatives on both sides with respect to $x^i$.
\section{Lower-order Hamiltonians}\label{app_Ham}

In this appendix we give the results for all the Hamiltonians given in eq.(\ref{eq_Ham_pp}) and eq.(\ref{eq_Ham_SO}). 

\subsection{Non-spinning sector up to 3PN}

\begin{align}
\widetilde{\mathcal{H}}_{\text{0PN}}= -\frac{1}{\widetilde{r}}+\frac{1}{2}\widetilde{p}_r^2+\widetilde{L}^2\left( \frac{1}{2 \widetilde{r}^2}\right)
\end{align}

\begin{align}
\widetilde{\mathcal{H}}_{\text{1PN}}&= 
\frac{1}{2 \widetilde{r}^2}
+\widetilde{p}_r^2 \left(-\frac{\nu }{\widetilde{r}}-\frac{3}{2 \widetilde{r}}\right)
+\widetilde{p}_r^4\left(\frac{3 \nu }{8}-\frac{1}{8}\right) \nonumber\\
&+\widetilde{L}^2 \left(-\frac{\nu }{2 \widetilde{r}^3}-\frac{3}{2 \widetilde{r}^3}+\widetilde{p}_r^2 \left(\frac{3 \nu }{4 \widetilde{r}^2}-\frac{1}{4 \widetilde{r}^2}\right)\right)
+\widetilde{L}^4 \left(\frac{3 \nu }{8 \widetilde{r}^4}-\frac{1}{8 \widetilde{r}^4}\right)
\end{align}

\begin{align}
\widetilde{\mathcal{H}}_{\text{2PN}}&=
-\frac{\nu }{4 \widetilde{r}^3}-\frac{1}{2 \widetilde{r}^3}
+\widetilde{p}_r^2 \left(\frac{9 \nu }{2 \widetilde{r}^2}+\frac{9}{4 \widetilde{r}^2}\right)
+\widetilde{p}_r^4 \left(-\frac{\nu ^2}{\widetilde{r}}-\frac{5 \nu }{2 \widetilde{r}}+\frac{5}{8 \widetilde{r}}\right)
+\widetilde{p}_r^6\left(\frac{5 \nu ^2}{16}-\frac{5 \nu }{16}+\frac{1}{16}\right) \nonumber\\
&+\widetilde{L}^2 \left(\frac{3 \nu }{\widetilde{r}^4}+\frac{11}{4 \widetilde{r}^4}+\widetilde{p}_r^2 \left(-\frac{\nu ^2}{\widetilde{r}^3}-\frac{9 \nu }{2 \widetilde{r}^3}+\frac{5}{4 \widetilde{r}^3}\right)+\widetilde{p}_r^4 \left(\frac{15 \nu ^2}{16 \widetilde{r}^2}-\frac{15 \nu }{16 \widetilde{r}^2}+\frac{3}{16 \widetilde{r}^2}\right)\right)\nonumber\\
&+\widetilde{L}^4 \left(-\frac{3 \nu ^2}{8 \widetilde{r}^5}-\frac{2 \nu }{\widetilde{r}^5}+\frac{5}{8 \widetilde{r}^5}+\widetilde{p}_r^2 \left(\frac{15 \nu ^2}{16 \widetilde{r}^4}-\frac{15 \nu }{16 \widetilde{r}^4}+\frac{3}{16 \widetilde{r}^4}\right)\right)
+\widetilde{L}^6 \left(\frac{5 \nu ^2}{16 \widetilde{r}^6}-\frac{5 \nu }{16 \widetilde{r}^6}+\frac{1}{16 \widetilde{r}^6}\right)
\end{align}

\begin{align}
\widetilde{\mathcal{H}}_{\text{3PN}}&=
\frac{\left(269-72 \pi ^2\right) \nu }{72 \widetilde{r}^4}+\frac{11}{8 \widetilde{r}^4}
+\widetilde{p}_r^2 \left(-\frac{5 \nu ^2}{2 \widetilde{r}^3}-\frac{\left(6368+207 \pi ^2\right) \nu }{288 \widetilde{r}^3}-\frac{3}{4 \widetilde{r}^3}\right)
+\widetilde{p}_r^4 \left(\frac{475 \nu ^2}{48 \widetilde{r}^2}+\frac{383 \nu }{48 \widetilde{r}^2}-\frac{25}{16 \widetilde{r}^2}\right)\nonumber\\
&+\widetilde{p}_r^6 \left(-\frac{\nu ^3}{\widetilde{r}}-\frac{3 \nu ^2}{\widetilde{r}}+\frac{21 \nu }{8 \widetilde{r}}-\frac{7}{16 \widetilde{r}}\right)
+\widetilde{p}_r^8\left(\frac{35 \nu ^3}{128}-\frac{35 \nu ^2}{64}+\frac{35 \nu }{128}-\frac{5}{128}\right) \nonumber\\
&+\widetilde{L}^2 \Bigg(-\frac{5 \nu ^2}{4 \widetilde{r}^5}+\frac{\left(207 \pi ^2-112\right) \nu }{576 \widetilde{r}^5}-\frac{21}{4 \widetilde{r}^5}+\widetilde{p}_r^2 \left(\frac{97 \nu ^2}{8 \widetilde{r}^4}+\frac{119 \nu }{8 \widetilde{r}^4}-\frac{27}{8 \widetilde{r}^4}\right)\nonumber \\
&\quad\quad\quad\quad+\widetilde{p}_r^4 \left(-\frac{3 \nu ^3}{2 \widetilde{r}^3}-\frac{8 \nu ^2}{\widetilde{r}^3}+\frac{59 \nu }{8 \widetilde{r}^3}-\frac{21}{16 \widetilde{r}^3}\right)+\widetilde{p}_r^6 \left(\frac{35 \nu ^3}{32 \widetilde{r}^2}-\frac{35 \nu ^2}{16 \widetilde{r}^2}+\frac{35 \nu }{32 \widetilde{r}^2}-\frac{5}{32 \widetilde{r}^2}\right)\Bigg)\nonumber\\
&+\widetilde{L}^4 \Bigg(+\frac{65 \nu ^2}{16 \widetilde{r}^6}+\frac{121 \nu }{16 \widetilde{r}^6}-\frac{29}{16 \widetilde{r}^6}+\widetilde{p}_r^2 \left(-\frac{9 \nu ^3}{8 \widetilde{r}^5}-\frac{115 \nu ^2}{16 \widetilde{r}^5}+\frac{55 \nu }{8 \widetilde{r}^5}-\frac{21}{16 \widetilde{r}^5}\right)\nonumber\\
&\quad\quad\quad\quad+\widetilde{p}_r^4 \left(\frac{105 \nu ^3}{64 \widetilde{r}^4}-\frac{105 \nu ^2}{32 \widetilde{r}^4}+\frac{105 \nu }{64 \widetilde{r}^4}-\frac{15}{64 \widetilde{r}^4}\right)\Bigg)\nonumber\\
&+\widetilde{L}^6 \left(-\frac{5 \nu ^3}{16 \widetilde{r}^7}-\frac{35 \nu ^2}{16 \widetilde{r}^7}+\frac{17 \nu }{8 \widetilde{r}^7}-\frac{7}{16 \widetilde{r}^7}+\widetilde{p}_r^2 \left(\frac{35 \nu ^3}{32 \widetilde{r}^6}-\frac{35 \nu ^2}{16 \widetilde{r}^6}+\frac{35 \nu }{32 \widetilde{r}^6}-\frac{5}{32 \widetilde{r}^6}\right)\right)\nonumber\\
&+\widetilde{L}^8 \left(\frac{35 \nu ^3}{128 \widetilde{r}^8}-\frac{35 \nu ^2}{64 \widetilde{r}^8}+\frac{35 \nu }{128 \widetilde{r}^8}-\frac{5}{128 \widetilde{r}^8}\right)
\end{align}

\subsection{Spin-orbit sector up to \NNLO}

\begin{align}
\widetilde{\mathcal{H}}_{\text{LO}}= 
\frac{\left(\widetilde{\textbf{S}}_{(1)}\cdot \widetilde{\textbf{L}}\right)}{q}  \left\{\frac{3 \nu}{2 \widetilde{r}^3}\right\}
+\left(\widetilde{\textbf{S}}_{(1)}\cdot \widetilde{\textbf{L}}\right)   \left\{\frac{2 \nu }{\widetilde{r}^3}\right\} + (1\leftrightarrow 2)
\end{align}

\begin{align}
\widetilde{\mathcal{H}}_{\text{NLO}}&= 
\frac{\left(\widetilde{\textbf{S}}_{(1)}\cdot \widetilde{\textbf{L}}\right)}{q}   \left\{-\frac{5 \nu ^2}{4 \widetilde{r}^4}-\frac{5 \nu }{\widetilde{r}^4}+\widetilde{p}_r^2 \left(\frac{17 \nu ^2}{4 \widetilde{r}^3}-\frac{5 \nu }{8 \widetilde{r}^3}\right)+
\widetilde{L}^2 \left(\frac{5 \nu ^2}{4 \widetilde{r}^5}-\frac{5 \nu }{8 \widetilde{r}^5}\right)
\right\}\nonumber\\
&+\left(\widetilde{\textbf{S}}_{(1)}\cdot \widetilde{\textbf{L}}\right)   \left\{-\frac{5 \nu ^2}{4 \widetilde{r}^4}-\frac{6 \nu }{\widetilde{r}^4}+\widetilde{p}_r^2 \left(\frac{43 \nu ^2 }{8 \widetilde{r}^3}\right)
+\widetilde{L}^2 \left(\frac{13  \nu ^2}{8 \widetilde{r}^5}\right)
\right\}\nonumber\\
& + (1\leftrightarrow 2)
\end{align}

\begin{align}
\widetilde{\mathcal{H}}_{\rm N^2LO}&= 
\frac{\left(\widetilde{\textbf{S}}_{(1)}\cdot \widetilde{\textbf{L}}\right)}{q} \Bigg\{
\frac{41 \nu ^2}{8 \widetilde{r}^5}+\frac{21 \nu }{2 \widetilde{r}^5}+\widetilde{p}_r^2 \left(-\frac{63 \nu ^3}{16 \widetilde{r}^4}-\frac{399 \nu ^2}{16 \widetilde{r}^4}+\frac{27 \nu }{8 \widetilde{r}^4}\right)+\widetilde{p}_r^4 \left(\frac{131 \nu ^3}{16 \widetilde{r}^3}-\frac{7 \nu ^2}{2 \widetilde{r}^3}+\frac{7 \nu }{16 \widetilde{r}^3}\right)\nonumber\\
&\quad\quad\quad\quad\quad+\widetilde{L}^2 \left(\widetilde{p}_r^2 \left(\frac{73 \nu ^3}{16 \widetilde{r}^5}-\frac{11 \nu ^2}{2 \widetilde{r}^5}+\frac{7 \nu }{8 \widetilde{r}^5}\right)-\frac{17 \nu ^3}{16 \widetilde{r}^6}-\frac{133 \nu ^2}{16 \widetilde{r}^6}+\frac{27 \nu }{8 \widetilde{r}^6}\right)+
\widetilde{L}^4 \left(\frac{17 \nu ^3}{16 \widetilde{r}^7}-\frac{2 \nu ^2}{\widetilde{r}^7}+\frac{7 \nu }{16 \widetilde{r}^7}\right)
\Bigg\}\nonumber\\
&+\left(\widetilde{\textbf{S}}_{(1)}\cdot \widetilde{\textbf{L}}\right) \Bigg\{
\frac{25 \nu ^2}{4 \widetilde{r}^5}+\frac{13 \nu }{\widetilde{r}^5}+\widetilde{p}_r^2 \left(-\frac{63 \nu ^3}{16 \widetilde{r}^4}-\frac{123 \nu ^2}{4 \widetilde{r}^4}+\frac{3 \nu }{\widetilde{r}^4}\right)+\widetilde{p}_r^4 \left(\frac{10 \nu ^3}{\widetilde{r}^3}-\frac{17 \nu ^2}{8 \widetilde{r}^3}\right)\nonumber\\
&\quad\quad\quad\quad\quad+\widetilde{L}^2 \left(\widetilde{p}_r^2 \left(\frac{23 \nu ^3}{4 \widetilde{r}^5}-\frac{19 \nu ^2}{8 \widetilde{r}^5}\right)-\frac{17 \nu ^3}{16 \widetilde{r}^6}-\frac{197 \nu ^2}{16 \widetilde{r}^6}-\frac{\nu }{\widetilde{r}^6}\right)+\widetilde{L}^4 \left(\frac{11 \nu ^3}{8 \widetilde{r}^7}-\frac{19 \nu ^2}{16 \widetilde{r}^7}\right)
\Bigg\}\nonumber\\
& + (1\leftrightarrow 2)
\end{align}

\bibliographystyle{JHEP}
\bibliography{biblio}

\end{document}